\documentclass[twocolumn,pre,amsmath,amssymb,showpacs]{revtex4-1}
\usepackage{graphicx,color}
\usepackage{amsmath,amssymb,latexsym,amsthm}
\usepackage{pgfplots}
\usepackage{mathtools}
\usepackage{mhchem}

\usepackage{tikz}
\usepgfplotslibrary{fillbetween}

\usetikzlibrary{chains,shapes,fit,calc,arrows}
\usepackage{pgfplots, pgfplotstable}
\usepgfplotslibrary{groupplots}
\usepackage{color}
\newcommand{\mean}[1]{{\left< #1 \right>}}

\usepackage{hyperref}
\hypersetup{%
    colorlinks=true, linktocpage=true, pdfstartpage=1, pdfstartview=FitV,%
    breaklinks=true, pdfpagemode=UseNone, pageanchor=true, pdfpagemode=UseOutlines,%
    plainpages=false, bookmarksnumbered, bookmarksopen=true, bookmarksopenlevel=1,%
    hypertexnames=true, pdfhighlight=/O,
    urlcolor=brown, linkcolor=red, citecolor=blue, 
    pdftitle={The dissipation-time uncertainty relation},%
    pdfauthor={Gianmaria Falasco, Massimiliano Esposito},%
    pdfsubject={},%
    pdfkeywords={},%
    pdfcreator={pdfLaTeX},%
    pdfproducer={LaTeX REVTeX}%
}

\begin{document}

\title{The dissipation-time uncertainty relation}
\author{Gianmaria Falasco}
\email{gianmaria.falasco@uni.lu}
\affiliation{Complex Systems and Statistical Mechanics, Department of Physics and Materials Science, University of Luxembourg, L-1511 Luxembourg, Luxembourg}
\author{Massimiliano Esposito}
\email{massimiliano.esposito@uni.lu}
\affiliation{Complex Systems and Statistical Mechanics, Department of Physics and Materials Science, University of Luxembourg, L-1511 Luxembourg, Luxembourg}
\date{\today}
\begin{abstract}

We show that the dissipation rate bounds the rate at which  physical processes can be performed in stochastic systems far from equilibrium. Namely, for rare processes we prove the fundamental tradeoff $\langle \dot S_\text{e} \rangle  \mathcal{T} \geq k_{\text{B}} $ between the entropy flow $\langle \dot S_\text{e} \rangle$ into the reservoirs and the mean time $\mathcal{T}$ to complete a process. This dissipation-time uncertainty relation is a novel form of speed limit: the smaller the dissipation, the larger the time to  perform a process.

\end{abstract}

\maketitle

Despite operating in noisy environments, complex systems are capable of actuating processes at finite precision and speed. 
Living systems in particular perform processes that are precise and fast enough to sustain, grow and replicate themselves. 
To this end, nonequilibrium conditions are required. 
Indeed, no process that is based on a continuous supply of (matter, energy, etc.) currents can take place without dissipation.

Recently, an intrinsic limitation on precision set by dissipation has been established by thermodynamic uncertainty relations \cite{bar15a, hor17, pro17, dec18, fal19, van20}. Roughly speaking, these inequalities state that the squared mean-to-variance ratio of currents is upper bonded by (a function of) the entropy production. Despite producing loose bounds for some specific models \cite{fal19n, mar19}, their fundamental importance is undeniable as they demonstrate that thermodynamics broadly constrains nonequilibrium dynamics \cite{hor19}.

For speed instead, an equivalent limitation set by dissipation can only be speculated.
For example, we know from macroscopic thermodynamics that thermodynamic machines will produce entropy to deliver finite power. 
Yet, a constraint on a par with thermodynamic uncertainty relations, only based on dissipation, is still lacking.

Efforts in this direction have appeared lately \cite{sha18, oku18}, inspired by research on quantum speed limits which are bounds on the time needed to transform a system from one state into another \cite{def17}. When extended to classical stochastic dynamics, these relations acquire a somewhat formal appearance \cite{shi18, ito18, ito18b, nic20}. In their most explicit form they bound the distance between an initial state and a final one at time $t$---technically, the 1-norm between the two probability distributions---in terms of the chosen time $t$, the dissipation, and other kinetic features of the system \cite{shi18}. However, many systems of interest, especially biological ones, operate under stationary (or time-periodic) conditions \cite{gne18}. They do not involve any (net) transformation in the system's state. The changes are confined to the reservoirs that fuel the nonequilibrium dynamics via mass or energy exchanges, for instance. Furthermore, the kinetic features of these systems are hardly known \cite{bat16, li19}. 

We show in this Letter that the dissipation alone suffices to bound the pace at which any stationary (or time-periodic) process can be performed. To do so, we set up the most appropriate framework to describe non-transient operations. Namely, we unambiguously define the process duration by the first-passage time for an observable to reach a given threshold \cite{gin17fp, gar17, ner17, ner20}. We first derive a bound  for the rate of the process $r$, uniquely specified by the survival probability that the process is not yet completed at time $t$ \cite{red01}. Then, for rare nonequilibrium processes that posses a constant rate, we obtain an uncertainty relation between the average duration of the process $\mathcal{T}=1/r$ and the mean dissipation rate in the reservoirs $\langle \dot S_\text{e} \rangle$. This novel speed limit applies to far from equilibrium system affected by weak fluctuations. The latter can arise in presence of weak noise \cite{gra87}, as found in transition state theory \cite{han95} or in macroscopic fluctuation theory \cite{ber15}, and when the processes is set by large thresholds \cite{gin17fp}. 

We start by considering stochastic trajectories $\omega_t$  of duration $t$---a list of states $x_{t'}$ with $ t' \in [0, t]$---in a space $\Omega_t$ with a stationary probability measure $P(\omega_t)= P(\omega_t|x_0) \rho(x_0)$, with $\rho(x_0)$ the stationary probability density of state $x_0$. 
We can think of $\omega_t$ as a diffusion or a jump process describing a nonequilibrium system subjected to the action of nonconservative forces which may be mechanical or generated by reservoirs with different temperature or chemical potentials, for instance.  
Also non-Markovian dynamics \cite{spe07, esp08} or unravelled quantum trajectories of open systems \cite{car19} may fit into the following framework. 
We introduce the stopping time $\tau:= \text{inf} \{t   \geq 0 : O(\omega_t) \in D\}$ as the minimum time for an observable $O: \Omega_t \to \mathbb{R}$ to reach values belonging to a specific domain $D \subset \mathbb{R}$, loosely referred to as `threshold'. The observable $O$, the threshold $D$, and the time $\tau$, define the physical process and its duration.
For concreteness, $\tau$ may be the minimum time to displace a mass, or to exchange a given amount of energy or particles with a reservoir. 
In general, it represents the time needed for a specific physical process to be carried out by the system. 

We next identify the space of `survived' trajectories at time $t$, $\Omega^{\text{s}}_t$, such that if $\omega_t \in \Omega^{\text{s}}_t$ then $O(\omega_t) \not \in D$. They correspond to trajectories in which the process is not completed. The associated probability that the process is not yet completed at time $t$ is expressed by the survival probability $p^\text{s}(t):= \text{Prob}(\tau > t) $, which satisfies $p^\text{s}(0)=1$ by definition. We can formally write it as 
\begin{align}\label{ps}
p^\text{s}(t) =  \sum_{\omega_t \in \Omega_t^{\text{s}}} P(\omega_t)=  \sum_{\omega_t \in \Omega_t } \chi(O(\omega_t)) P(\omega_t)
\end{align}
where $\chi(O(\omega_t))$ equals 1 if $O(\omega_t) \not \in D$ and zero otherwise.

We then consider the (involutive) transformation $\omega_t \mapsto  \tilde {\omega_t}$ that time-reverses the order of the states $x_{t'}$ (possibly changing sign, according to their parity). 
This allows to define the log ratio
\begin{align}\label{Sigma}
k_{\text{B}} \log \frac{P(\omega_t)}{P(\tilde {\omega_t})} =: \int_0^t dt' \dot \Sigma(\omega_t) .
\end{align} 
We note that $P(\tilde {\omega_t})=P(\tilde {\omega_t}|\tilde x_0) \rho(\tilde x_0) $ is the probability measure of time-reversed trajectories evolving with the original dynamics and starting from the stationary probability distribution $\rho(\tilde x_0)$.
If the dynamics obeys local detailed balance \cite{sei12, van15}, $\dot \Sigma= \dot S_\text{e} + dS/dt' $ is the entropy production rate at time $t'$, which splits into the entropy flux in the reservoirs, $\dot S_\text{e}$, plus the time derivative of the system entropy  $S = -\log\rho(x_{t'})$. 

Applying the time-reversal to \eqref{ps} and using \eqref{Sigma} we find
\begin{align}
p^s(t)&= \sum_{\omega_t \in \Omega_t } \chi(O(\tilde \omega_t))e^{-\int_0^t dt' \dot \Sigma(\omega_t)/k_\text{B} } P( \omega_t) \label{meantilde} 
\end{align}
after the relabeling $\tilde \omega_t \to \omega_t$. Notice that the sum in \eqref{meantilde} is restricted by $\chi(O(\tilde \omega_t))$ to a subset of trajectories $\tilde \Omega^{\text{s}}_t$ which differs from $ \Omega^{\text{s}}_t$ if $\tilde O(\omega_t):= O( \tilde  \omega_t) \neq O(\omega_t)$. This defines a different process, named reversed process, whose associated survival probability is $\tilde p^\text{s}(t) :=  \sum_{\omega_t \in \Omega_t } \chi(\tilde O(\omega_t)) P( \omega_t)$. Hence, we arrive at the modified integral fluctuation relation
\begin{align}
p^\text{s}(t) 
&=\tilde p^\text{s}(t) \mean{e^{-\int_0^t dt' \dot \Sigma/k_{\text{B}}}}_{\tilde{\mathrm{s}}}.\label{ftopen}
\end{align}
Hereafter, $\mean{F}_{\tilde{\mathrm{s}}}:= \sum_{\omega_t \in \tilde \Omega_t^\text{s}} F(\omega_t)  P( \omega_t)/ \tilde p^\text{s}(t)$ denotes the normalized average of the generic observable $F$ on the set of survived trajectories $\tilde \Omega_t^{\text{s}}$.
One should note that (\ref{ftopen}) appears in implicit form in Ref. \cite{mur14, fun15, mur18} as a generalized fluctuation theorem holding when a subset of forward trajectories have no time-reversal equivalent \cite{supinf}. Our crucial new ingredient is to define $\Omega^\text{s}$ and $\tilde \Omega^\text{s}$ via the choice of an observable and a threshold, and to assign stopping times to trajectories in that subset.

As a consequence of Jensen's inequality, \eqref{ftopen} yields 
\begin{align}\label{psbound}
p^\text{s}(t) \geq  \tilde p^s(t) e^{-\int_0^t dt'   \langle \dot  \Sigma \rangle_{\tilde{\mathrm{s}}}  /k_{\text{B}}} ,
\end{align}
which gives a bound on the pace at which the two processes proceed. 
Since survival probabilities are positive and monotonically decreasing, one can define the instantaneous rate $r(t)$ of the process as \cite{for11}
\begin{align}\label{rate}
r(t):=- \frac{1}{ p^\text{s}(t)} \frac{d p^\text{s}}{d t}(t) ,
\end{align} 
and analogously for $\tilde r(t)$.
Because \eqref{psbound} holds for any positive time $t'$, we find that the entropy production rate bounds the difference of the process rates
\begin{align}\label{rbound}
\frac{1}{k_{\text{B}}}\langle \dot \Sigma \rangle_{\tilde{\mathrm{s}}}(t')  \geq r(t')-\tilde r(t') .
 \end{align}

When the mean time of the process exists, i.e. when $\mean{\tau}=\int_0^{\infty} dt \, p^{\text{s}}(t)=:\mathcal{T}$ is finite \cite{red01}, \eqref{rbound} can be turned into a bound on $\mathcal{T}$. First, we assume that both processes are rare so that their survival probabilities read $p^\text{s}(t)= e^{-rt}$ and $\tilde p^\text{s}(t)= e^{-\tilde rt}$, which implies $\mathcal{T}=1/r$. Second, we choose the threshold $D$ such that $\tilde r \ll r$. This implies that $\tilde p^\text{s}(t') =1$ for all times $t'  \ll1 /\tilde r$ and the entropy production rate coincides with the (constant) mean entropy flux of the stationary dynamics, $\langle \dot \Sigma \rangle_{\tilde{\textrm{s}}}=\langle \dot S_\text{e} \rangle $. 
Under these assumptions, \eqref{rbound} simplifies to our main result: the dissipation-time uncertainty principle
\begin{align}\label{speeddiss}
 \langle \dot S_\text{e} \rangle \, \mathcal{T} \geq  k_{\text{B}}.
\end{align}
The fundamental implication of this result is that to realize a rare nonequilibrium process in a given (average) time $\mathcal{T}$ at least $k_{\text{B}}/\mathcal{T}$ must be dissipated in the reservoirs. Rare processes are effectively Poissonian processes on long timescales compared to the fast microscopic dynamics. They arise in two broad classes of problems. First in the presence of weak noise, which is typically the case for problems described by transition state theory, or by macroscopic fluctuation theory where fluctuations are exponentially suppressed in the system size. Second, for large thresholds, i.e. when the domain $D$ can be reached only by atypical fluctuations.  We will illustrate these two cases on paradigmatic models. 

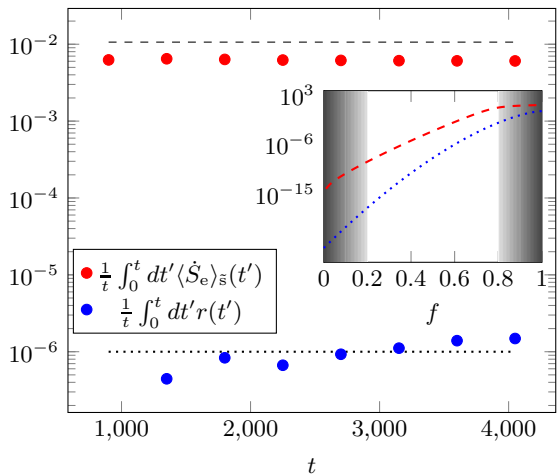
\begin{figure}[t]
	\centering
	\begin{tikzpicture}
	\begin{axis}[width=0.45\textwidth, xlabel={$t$}, ymode= log, xlabel near ticks, ylabel near ticks, ylabel shift = -4 pt,legend style={at={(0.01,0.3)},anchor=west}] 
	\addplot[color=red, only marks] coordinates{(900, 0.00624902)(1350, 0.00646795)(1800, 0.00634419)(2250, 0.00622373)(2700, 0.00616055)(3150, 0.00610911)(3600, 0.00607739)(4050, 0.00606302)};
	\addlegendentry{$\frac 1 t \int_0^t dt' \langle \dot S_\text{e} \rangle_{\tilde{ \text{s}}}(t')$}
\addplot[color=blue, only marks] coordinates{(1350,  4.44467*10^-07)(1800, 8.33959*10^-7)(2250,6.67167*10^-7)(2700, 9.27085*10^-7)(3150, 1.11306*10^-6)(3600, 1.39237*10^-6)(4050, 1.48594*10^-6)};
\addlegendentry{$\frac 1 t  \int_0^t dt' r(t')$}
\addplot[color=black, no markers,style={thick}, dotted] coordinates{(900,10^-6)(4050,10^-6)} ;
\addplot[color=black, no markers, dashed] coordinates{(900,0.0106505)(4050,0.0106505)} ;
	\end{axis}
	\begin{axis}[anchor=south west, xshift=3.4cm, yshift=2cm,width=0.25\textwidth,ymode= log, xlabel=$f$, xmin=0, xmax=1,ylabel near ticks, yticklabel pos=left, xlabel near ticks,ylabel style={rotate=-90}, ytick={0.000000000000001,0.000001,1000}]
	 \fill[shade, left color=black!70, right color=black!0 , fill opacity=1]
        (rel axis cs:0,0.003)--(rel axis cs:0.2,0.003)--
        (rel axis cs:0.2,0.997)--(rel axis cs:0,0.997)--cycle;
	 \fill[shade, left color=black!0, right color=black!70 , fill opacity=1]
        (rel axis cs:0.8,0.003)--(rel axis cs:1,0.003)--
        (rel axis cs:1,0.997)--(rel axis cs:0.8,0.997)--cycle;
\addplot[color=red, no markers,style={thick},dashed] coordinates{
(0.01,8.31775*10^-15)(0.02,2.36924*10^-14)(0.03,6.22419*10^-14)(0.04,1.35439*10^-13)(0.05,2.68862*10^-13)(0.06,5.05792*10^-13)(0.07,9.18987*10^-13)(0.08,1.60725*10^-12)(0.09,2.80681*10^-12)(0.1,4.81757*10^-12)(0.11,8.18725*10^-12)(0.12,1.37649*10^-11)(0.13,2.29345*10^-11)(0.14,3.79802*10^-11)(0.15,6.23296*10^-11)(0.16,1.01693*10^-10)(0.17,1.65035*10^-10)(0.18,2.6653*10^-10)(0.19,4.2851*10^-10)(0.2,6.86041*10^-10)(0.21,1.09401*10^-9)(0.22,1.73809*10^-9)(0.23,2.75157*10^-9)(0.24,4.34126*10^-9)(0.25,6.82719*10^-9)(0.26,1.07032*10^-8)(0.27,1.67291*10^-8)(0.28,2.60735*10^-8)(0.29,4.05173*10^-8)(0.3,6.27932*10^-8)(0.31,9.7053*10^-8)(0.32,1.49619*10^-7)(0.33,2.30065*10^-7)(0.34,3.52877*10^-7)(0.35,5.39914*10^-7)(0.36,8.24084*10^-7)(0.37,1.25482*10^-6)(0.38,1.90618*10^-6)(0.39,2.88893*10^-6)(0.4,4.36825*10^-6)(0.41,6.59002*10^-6)(0.42,9.91934*10^-6)(0.43,0.0000148971)(0.44,0.000022323)(0.45,0.0000333761)(0.46,0.0000497919)(0.47,0.0000741178)(0.48,0.000110085)(0.49,0.000163147)(0.5,0.000241252)(0.51,0.000355963)(0.52,0.000524055)(0.53,0.000769805)(0.54,0.00112826)(0.55,0.0016499)(0.56,0.00240718)(0.57,0.00350388)(0.58,0.00508814)(0.59,0.00737073)(0.6,0.0106505)(0.61,0.0153495)(0.62,0.0220608)(0.63,0.0316141)(0.64,0.0451617)(0.65,0.0642928)(0.66,0.0911779)(0.67,0.128746)(0.68,0.180889)(0.69,0.252679)(0.7,0.350549)(0.71,0.482384)(0.72,0.657388)(0.73,0.885603)(0.74,1.17692)(0.75,1.53954)(0.76,1.97799)(0.77,2.49113)(0.78,3.07094)(0.79,3.70259)(0.8,4.36632)(0.81,5.0405)(0.82,5.70501)(0.83,6.34393)(0.84,6.94677)(0.85,7.50838)(0.86,8.02804)(0.87,8.50808)(0.88,8.95261)(0.89,9.36657)(0.9,9.75495)(0.91,10.1225)(0.92,10.4733)(0.93,10.811)(0.94,11.1386)(0.95,11.4586)(0.96,11.773)(0.97,12.0834)(0.98,12.3911)(0.99,12.6972)(1.,13.0026)
};
\addplot[color=blue, no markers,style={thick},dotted] coordinates{ 
(0.,1.52466*10^-25)(0.01,3.73038*10^-25)(0.02,9.07512*10^-25)(0.03,2.19518*10^-24)(0.04,5.27963*10^-24)(0.05,1.26257*10^-23)(0.06,3.00206*10^-23)(0.07,7.0974*10^-23)(0.08,1.66836*10^-22)(0.09,3.89935*10^-22)(0.1,9.06156*10^-22)(0.11,2.09372*10^-21)(0.12,4.80992*10^-21)(0.13,1.09864*10^-20)(0.14,2.49502*10^-20)(0.15,5.63357*10^-20)(0.16,1.26469*10^-19)(0.17,2.82273*10^-19)(0.18,6.26378*10^-19)(0.19,1.38191*10^-18)(0.2,3.03108*10^-18)(0.21,6.60969*10^-18)(0.22,1.43294*10^-17)(0.23,3.08836*10^-17)(0.24,6.61729*10^-17)(0.25,1.40953*10^-16)(0.26,2.98474*10^-16)(0.27,6.28302*10^-16)(0.28,1.31478*10^-15)(0.29,2.73497*10^-15)(0.3,5.65533*10^-15)(0.31,1.16242*10^-14)(0.32,2.37496*10^-14)(0.33,4.82316*10^-14)(0.34,9.73594*10^-14)(0.35,1.95337*10^-13)(0.36,3.89531*10^-13)(0.37,7.7204*10^-13)(0.38,1.52078*10^-12)(0.39,2.97721*10^-12)(0.4,5.79239*10^-12)(0.41,1.11995*10^-11)(0.42,2.15188*10^-11)(0.43,4.10868*10^-11)(0.44,7.79539*10^-11)(0.45,1.46964*10^-10)(0.46,2.75298*10^-10)(0.47,5.12392*10^-10)(0.48,9.47521*10^-10)(0.49,1.74079*10^-9)(0.5,3.17729*10^-9)(0.51,5.76106*10^-9)(0.52,1.03768*10^-8)(0.53,1.8566*10^-8)(0.54,3.29948*10^-8)(0.55,5.82404*10^-8)(0.56,1.02102*10^-7)(0.57,1.77764*10^-7)(0.58,3.07352*10^-7)(0.59,5.27694*10^-7)(0.6,8.99608*10^-7)(0.61,1.52273*10^-6)(0.62,2.55894*10^-6)(0.63,4.26908*10^-6)(0.64,7.06989*10^-6)(0.65,0.0000116215)(0.66,0.0000189602)(0.67,0.0000306989)(0.68,0.0000493241)(0.69,0.0000786341)(0.7,0.000124375)(0.71,0.000195155)(0.72,0.000303741)(0.73,0.000468867)(0.74,0.000717736)(0.75,0.00108941)(0.76,0.00163932)(0.77,0.00244521)(0.78,0.00361477)(0.79,0.00529516)(0.8,0.00768472)(0.81,0.0110469)(0.82,0.0157261)(0.83,0.0221649)(0.84,0.0309214)(0.85,0.0426852)(0.86,0.0582886)(0.87,0.0787091)(0.88,0.105059)(0.89,0.138551)(0.9,0.180445)(0.91,0.231943)(0.92,0.294055)(0.93,0.367399)(0.94,0.451946)(0.95,0.546699)(0.96,0.649296)(0.97,0.755509)(0.98,0.85853)(0.99,0.947525)(1.,1.)
};
\end{axis}
	\end{tikzpicture}
	\caption{Speed limit for the dynamics \eqref{langev} with process \eqref{process}, obtained by numerical averages over $2 \cdot 10^3$ trajectories with $a=1$, $\beta^{-1}=0.07, f=0.6, N=2$, $k_{\text{B}}=1$. Dashed and dotted lines correspond to the analytical estimation of time-integrated $\langle \dot S_\text{e} \rangle$ and $r$, respectively, while the dots denote their numerical values. At short times $t \lesssim 10^3$ the numerical estimation of $r(t)$ is impeded by the finite statistics. \emph{Inset: } The analytical approximation of  $\langle \dot S_\text{e} \rangle$ (dashed) and $r$ (dotted) as function of the tilt $f$. }
	\label{fig:lang}
\end{figure}

The first example represents overdamped particle transport in one spatial dimension.
The dynamics follows the Langevin equation 
\begin{align}\label{langev}
\dot x= -U'(x) + \sqrt{2/\beta} \xi
\end{align}
with periodic boundary conditions in $x \in [0, 2 \pi]$. 
Here, $(k_{\text{B}} \beta)^{-1}$ denotes temperature, $\xi$ is a zero-mean Gaussian white noise of unit variance, and $U(x)=a \cos(x)-fx$ is a periodic potential superimposed to a constant nonconservative tilt $f>0$. 
This model describes a wealth of transport processes ranging from loaded molecular motors \cite{jul97} to electrons across Josephson junctions \cite{die80}.
For weak noises, $\eqref{speeddiss}$ applies in the stationary regime for the process of transporting the coordinate $x$ over $N >0$ periods, i.e. 
\begin{align}\label{process}
O= \int_0^t \dot  x_{t'} d t', \quad D=[2 \pi N, \infty),
\end{align}
Indeed, when $\beta^{-1}$ is small with respect to the smaller energy barrier $\Delta U_\text{min}$, escaping from the tilted potential well is a weak noise problem \cite{bou16}. Namely, the process \eqref{process} unfolds with a single rate, which can be roughly estimated as the product of $N$ Arrhenius factors, i.e. $r \sim e^{-N \beta \Delta U_\text{min} }$  (see Fig.~\ref{fig:lang}). Similarly, the reversed process defined by the trajectories realizing a negative current $ \tilde O =  - 2 \pi N$ takes place with rate $\tilde r \sim e^{-N \beta \Delta U_\text{max} } \ll r $ (where $U_\text{max}$ is the larger energy barrier), which is negligible for sufficiently large $f$. Thus, the uncertainty principle \eqref{speeddiss} applies with the stationary entropy flow given by
\begin{align}\label{Sediff}
\langle \dot S_\text{e} \rangle= k_{\text{B}} f\beta \left[f\!+\!  a\! \int dx \rho(x) \sin (x) \right].
\end{align}
Here $ \rho$ is the stationary probability distribution associated to \eqref{langev}, which  is well approximated by the local-equilibrium distribution $\rho(x) \sim e^{-\beta U(x)}$  (see Fig.~\ref{fig:lang}) \cite{ris89}. 

Note however that these conditions on $r$ and  $\langle \dot S_\text{e} \rangle$ break down in two limits, when $f \to 1$ and when $f \to 0$. 
When $f \to 1$, i.e. the value for which $\Delta U_\text{min}  \to 0$, the weak noise assumption becomes invalid, so the process is no longer rare. 
When $f \to 0$, i.e. close to detailed balance dynamics, the reverse process is no longer negligible, so that \eqref{rbound} cannot be simplified to \eqref{speeddiss}. Nonetheless, within its range of validity, the analytical estimation shows that \eqref{speeddiss} becomes tighter as the distance from equilibrium increases (see inset in Fig.~\ref{fig:lang}).

\begin{figure}[t]
	\centering
	\begin{tikzpicture}
	\begin{axis}[width=0.45\textwidth, xlabel={$t$}, ymode= log, xlabel near ticks, ylabel near ticks, ylabel shift = -4 pt,legend style={at={(0.03,0.25)},anchor=west}] 
	\addplot[color=red, only marks] coordinates{
	(25 ,0.401336)
(50, 0.381724)
(75 ,0.375662)
(100, 0.376226)
(125 ,0.374995)
(150 ,0.374991)
(175, 0.374469)
(200 ,0.372795)
(225 ,0.372138)
(250 ,0.371784)
(275 ,0.373523)
(300 ,0.372124)
(325 ,0.370685)
(350 ,0.370555)
(375 ,0.371451)
(400 ,0.369668)
(425 ,0.371813)
(450 ,0.369949)
(475 ,0.369502)
	};
	\addlegendentry{$\frac 1 t \int_0^t dt' \langle \dot S_\text{e} \rangle_{\tilde{\text{s}}}(t')$}
\addplot[color=blue, only marks,] coordinates{
(25 ,0.00121643)
(50 ,0.00146965)
(75 ,0.00148996)
(100, 0.00152942)
(125 ,0.00150404)
(150 ,0.00152129)
(175 ,0.00153358)
(200 ,0.00157378)
(225 ,0.0015816)
(250 ,0.00157867)
(275 ,0.00157391)
(300 ,0.00156986)
(325 ,0.00158572)
(350 ,0.00159099)
(375 ,0.00157779)
(400 ,0.00157843)
(425 ,0.00156995)
(450 ,0.00158269)
(475 ,0.00158191)
};
\addlegendentry{$\frac 1 t  \int_0^t dt' r(t')$}
\addplot[color=black, no markers,style={thick}, dashed] coordinates {(25,0.3678934885024677)(475,0.3678934885024677)};
	\end{axis}
	\begin{axis}[anchor=south west, xmin=1, xmax=10, xshift=3.4cm, yshift=2cm,width=0.25\textwidth,ymode= log, xlabel=$\Delta \beta$, ylabel near ticks, yticklabel pos=left, xlabel near ticks,ylabel style={rotate=-90}, ytick={1,0.1,0.01}]
\addplot[color=red, thick, dashed, domain=1:10] (x, {(exp ((0.5*x)/(1 + x - sqrt(1 + x*x))) - 
   exp ((0.5*x)/(-1 + x + sqrt(1 + x*x))))/( 1 + exp (-((0.5 *x*x)/(1 - sqrt(1 + x*x)))))*x } );
\addplot[color=red, only marks] coordinates{
(1/0.4 - 1/1.6, 0.695)(1/0.5 - 1/1.5, 0.373)(1/0.35 - 1/1.65,  0.96)(1/0.3- 1/1.7, 1.34) (1/0.2-1/1.8,2.8) (1/0.15-1/1.85, 4.33)(1/0.1-1/1.9, 7.2)
};
\addplot[color=blue,only marks] coordinates{ 
(1/0.5 -  1/1.5, 0.00157)(1/0.4 - 1/1.6,0.00326)(1/0.35 - 1/1.65, 0.00485)(1/0.3- 1/1.7, 0.0074) (1/0.2-1/1.8, 0.017) (1/0.15-1/1.85, 0.025)(1/0.1-1/1.9, 0.03)
};
\end{axis}
	\end{tikzpicture}
	\caption{Speed limit for the two-state system of the main text with process \eqref{process2}, obtained by numerical averages over $10^5$ Gillespie trajectories with $1/\beta_\text{h}=1.5, 1/\beta_\text{c}=0.5, \epsilon_2=1, \epsilon_1=0, E= 5, \delta=6, k_{\text{B}}=1$. The dashed line corresponds to $\langle \dot S_\text{e} \rangle$. At short times $t \lesssim 50$ the numerical estimation of $r(t)$ is impeded by the finite statistics. \emph{Inset:} time-averaged value of $\langle \dot S_\text{e} \rangle_{\tilde{\text{s}}}$ and process rate $r$ as function of $\Delta \beta := \beta_\text{c}-\beta_\text{h}$ at fixed average temperature $(1/\beta_\text{c}+1/\beta_\text{h})/2=1$. }
	\label{fig:twost}
\end{figure}
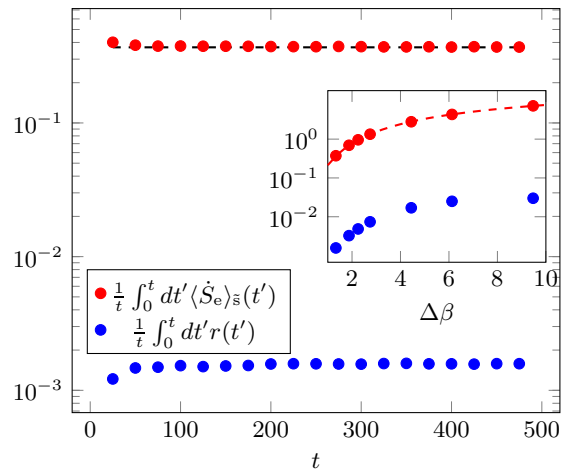

The second example represents energy transfer between two heat baths (at inverse temperatures $k_{\text{B}}\beta_\text{h}$ and $k_{\text{B}}\beta_\text{c}$, respectively) mediated by a two-level system. The latter performs Markovian jumps (corresponding to Poisson processes $dN^\nu_{i \to j}$) between the two states $i=\{1,2\}$ of energy $\epsilon_i$ with rates $w^{\nu}_{i \to j}= e^{-\beta_\nu(\epsilon_j-\epsilon_i)/2}$ associated to the baths $\nu=\{\text{h},\text{c}\}$. We define the process as the transfer of an energy $E$ in a fixed time $\delta$ into the cold bath $\nu=\text{c}$:
 \begin{align}\label{process2}
 O= (\epsilon_2-\epsilon_1) \int_0^{\delta} dt' \left[\frac{dN_{2 \to 1}^\text{c}}{dt'}-\frac{dN_{1 \to 2}^\text{c}}{dt'} \right], \,  D=[E, \infty),
 \end{align}
One may think of $E$ as an activation energy (e.g., of reaction \cite{han90}) and of $\delta$ as the timescale over which it may be dissipated.
For large $\delta^{-1}$ and/or $E>0$ (with respect to the rates $w^{\nu}_{i \to j}$ and the energy gap  $\Delta \epsilon:=|\epsilon_1-\epsilon_2|$, respectively) the process is realized by large fluctuations and, therefore, is rare. The rate of the reverse process, defined by extracting an energy larger than $E$ in the time $\delta$ from the cold reservoir, is in comparison negligible. Since the rate $r$ and the entropy flow $\langle \dot S_\text{e} \rangle_{\tilde{\text{s}}}\simeq \langle \dot S_\text{e} \rangle = k_{\text{B}} (1/\beta_\text{c}-1/\beta_\text{h}) (e^{\frac{ \Delta \epsilon \beta_\text{h} }{2}}-e^{\frac{\Delta \epsilon \beta_\text{c}}{2}})(e^{\frac{\Delta \epsilon \beta_\text{h} }{2}+\frac{\Delta \epsilon \beta_\text{c}}{2}}+1)^{-1}$  are constant, as shown in Fig.~\ref{fig:twost}, equation \eqref{speeddiss} holds true.

These two examples suggest that currents are natural observables to which our theory applies. For integrated currents, i.e. $\tilde O( \omega_t) =- O(\omega_t)$, our results can also be directly derived from the steady state fluctuation theorem \cite{supinf}. Note that in a similar setting, i.e. a large threshold for $O$, a complementary result based on large deviations theory is known, which holds when $O$ is an integrated current  \cite{gin17fp} (resp. counting observable \cite{gar17}) scaling linearly with $t$. It \emph{lower} bounds $\mathcal{T}^2$ by the variance of $\tau$ times the entropy production (resp. the dynamical activity). Our theory, instead, provides an upper bound only based on entropy production and allows to consider more general observables $O$, e.~g., that are not extensive in the trajectory duration $t$. However, our theory does not impose any constraint on processes defined by time symmetric observables, i.e. $\tilde O( \omega_t) = O(\omega_t)$. For them, $p^\text{s}$ and $\tilde p^\text{s}$ coincide so that \eqref{ftopen}  and \eqref{psbound} only give (for survived trajectories) the usual integral fluctuation theorem and the positivity of entropy production, respectively.

We then come back to the general theory, seeking extensions of the inequalities \eqref{rbound} and \eqref{speeddiss} beyond stationarity. Systems subject to time-dependent driving can be treated with a slight modification of the above derivation which includes time-reversal of the driving protocols $\lambda(t') \mapsto \lambda(t-t')$ both in the dynamics and in the initial probability of time-reversed trajectories \cite{rao18}, which we take as the periodic steady state $\rho_{\lambda(t)}(x_t)$ at the final value of the protocol $\lambda(t)$. Then,~\eqref{rbound} generalizes to 
\begin{align}\label{rbound2}
\frac{1}{k_{\text{B}}}\langle \dot \Sigma \rangle^\text{B}_{\tilde{\mathrm{s}}}(t')  \geq r^\text{F}(t')-\tilde r^\text{B}(t') ,
 \end{align}
 which bounds the difference of the rates of the forward and backward process with the dissipation in the backward dynamics. Importantly, \eqref{rbound} is recovered for time-symmetric driving protocols for which forward and backward dynamics coincide. Moreover, rare processes with sufficiently fast driving of small amplitude are still characterized (for $t'$ much larger than the driving period $t_\text{d}$)  only by the constant time-averaged rate $\bar r:=\frac{1}{t_\text{d}} \int_0^{t_\text{d}}dt'  r(t') \simeq \mathcal{T}^{-1}$, so that $p^{\text{s}}(t) \simeq e^{-t \bar r}$, and by the entropy flow rate $\overline {\langle \dot S_\text{e} \rangle}= \frac{1}{t_\text{d}} \int_0^{t_\text{d}}dt' \langle \dot S_\text{e}\rangle(t')$. For such processes, \eqref{speeddiss} holds in the integrated form 
\begin{align}
\overline {\langle \dot S_\text{e} \rangle} \, \mathcal{T} \geq  k_{\text{B}},
\end{align}
if the threshold $D$ is chosen such that the rate of the reversed process is negligible.
Our approach still holds in the most general case of driven transient dynamics with arbitrary threshold. Eqs. \eqref{psbound} and \eqref{rbound2} remain valid but their use can be impractical. On one hand, for processes with a survival probability decaying as a power-law
 $p^\text{s}(t) \sim 1/t^\gamma$, the definition \eqref{rate} is rather {\it ad hoc} since no characteristic timescale exists (the average time $\mathcal{T}$ even diverges for $\gamma \leq 1$). On the other hand, the entropy production rate is hard to estimate, being dependent on the survival probability itself. This clarifies how our results are complementary to the other classical speed limits, since our formalism is specifically tailored to different processes, which are marked by transformations in the reservoirs instead of in the system.

Our main result, \eqref{speeddiss}, asserts that a large dissipation allows for a fast process.
But this does not imply that increasing dissipation will necessarily speed up the process.  As for thermodynamic uncertainty relations \cite{fal19n, mar19}, kinetics aspects of the dynamics are essential to determine the tightness of the bound \cite{dit18}. Concerning tightness, it is known from the theory of deterministic dynamical systems that escape rate is proportional to the dissipation rate in open Hamiltonian systems where particle leakage on large scales is compatible with a drifted diffusive process \cite{bre96, gas07}. This result can be obtained within our framework considering the dynamics \eqref{langev} with $a=0$
and the process of transporting the coordinate $x$ to either $0$ or $L$. Since $\Omega^\text{s}=\tilde \Omega^\text{s}$ for such process, \eqref{speeddiss} does not directly apply but it can be easily generalized by using an appropriate auxiliary dynamics in \eqref{meantilde}, which is different from time reversal \cite{supinf}. Actually, this general strategy can be applied to derive system-specific bounds for processes that break symmetries other than time-reversal (such as reflection).

In summary, resorting to the concept of first passage time we have defined the physical process that a stochastic system can perform. 
This allowed us to describe stationary (resp., periodic) processes, that do not involve (resp., any net) transformation of states, and to define unambiguously the pace at which the process advances.
Irrespective of the stochastic dynamics, we have provided an upper bound on the rate at which a process is performed. This result becomes particularly useful for rare processes---e.g. realized by weak fluctuations or defined by large thresholds---far from equilibrium, where the bound reduces to the entropy flow in the reservoirs. This result suggests that the connection between our framework and the escape-rate theory of deterministic systems \cite{gas05} should be deepened, in particular for hyperbolic ones where the role of weak noise is taken up by chaos to produce a constant escape rate $r$. Furthermore, it calls for an extension of stochastic thermodynamics to systems with escape, given the recent work on the second law at stopping times \cite{ner20}, and the renewed fundamental interest in unstable dynamics \cite{vsi18, orn18} .

Finally, the dissipation-time uncertainty relation, together with the recent thermodynamic uncertainty relations show manifestly how thermodynamics constrains nonequilibrium dynamics. It also hints at a general emerging tradeoff between speed, precision, accuracy and dissipation \cite{meh12, dia13, sar14, ian16}, in which the role of information \cite{hor14, par15, kol18, deff20} only awaits to be explicitly uncovered.

This research was funded by  the European Research Council project NanoThermo (ERC-2015-CoG Agreement No. 681456). We thank Ptaszy\'{n}ski Krzysztof for valuable comments on an earlier draft of this work.

\bibliography{RefDFTUR}

\begin{thebibliography}{58}%
\makeatletter
\providecommand \@ifxundefined [1]{%
 \@ifx{#1\undefined}
}%
\providecommand \@ifnum [1]{%
 \ifnum #1\expandafter \@firstoftwo
 \else \expandafter \@secondoftwo
 \fi
}%
\providecommand \@ifx [1]{%
 \ifx #1\expandafter \@firstoftwo
 \else \expandafter \@secondoftwo
 \fi
}%
\providecommand \natexlab [1]{#1}%
\providecommand \enquote  [1]{``#1''}%
\providecommand \bibnamefont  [1]{#1}%
\providecommand \bibfnamefont [1]{#1}%
\providecommand \citenamefont [1]{#1}%
\providecommand \href@noop [0]{\@secondoftwo}%
\providecommand \href [0]{\begingroup \@sanitize@url \@href}%
\providecommand \@href[1]{\@@startlink{#1}\@@href}%
\providecommand \@@href[1]{\endgroup#1\@@endlink}%
\providecommand \@sanitize@url [0]{\catcode `\\12\catcode `\$12\catcode
  `\&12\catcode `\#12\catcode `\^12\catcode `\_12\catcode `\%12\relax}%
\providecommand \@@startlink[1]{}%
\providecommand \@@endlink[0]{}%
\providecommand \url  [0]{\begingroup\@sanitize@url \@url }%
\providecommand \@url [1]{\endgroup\@href {#1}{\urlprefix }}%
\providecommand \urlprefix  [0]{URL }%
\providecommand \Eprint [0]{\href }%
\providecommand \doibase [0]{http://dx.doi.org/}%
\providecommand \selectlanguage [0]{\@gobble}%
\providecommand \bibinfo  [0]{\@secondoftwo}%
\providecommand \bibfield  [0]{\@secondoftwo}%
\providecommand \translation [1]{[#1]}%
\providecommand \BibitemOpen [0]{}%
\providecommand \bibitemStop [0]{}%
\providecommand \bibitemNoStop [0]{.\EOS\space}%
\providecommand \EOS [0]{\spacefactor3000\relax}%
\providecommand \BibitemShut  [1]{\csname bibitem#1\endcsname}%
\let\auto@bib@innerbib\@empty
\bibitem [{\citenamefont {Barato}\ and\ \citenamefont
  {Seifert}(2015)}]{bar15a}%
  \BibitemOpen
  \bibfield  {author} {\bibinfo {author} {\bibfnamefont {A.~C.}\ \bibnamefont
  {Barato}}\ and\ \bibinfo {author} {\bibfnamefont {U.}~\bibnamefont
  {Seifert}},\ }\href@noop {} {\bibfield  {journal} {\bibinfo  {journal} {Phys.
  Rev. Lett.}\ }\textbf {\bibinfo {volume} {114}},\ \bibinfo {pages} {158101}
  (\bibinfo {year} {2015})}\BibitemShut {NoStop}%
\bibitem [{\citenamefont {Horowitz}\ and\ \citenamefont
  {Gingrich}(2017)}]{hor17}%
  \BibitemOpen
  \bibfield  {author} {\bibinfo {author} {\bibfnamefont {J.~M.}\ \bibnamefont
  {Horowitz}}\ and\ \bibinfo {author} {\bibfnamefont {T.~R.}\ \bibnamefont
  {Gingrich}},\ }\href@noop {} {\bibfield  {journal} {\bibinfo  {journal}
  {Phys. Rev. E}\ }\textbf {\bibinfo {volume} {96}},\ \bibinfo {pages} {020103}
  (\bibinfo {year} {2017})}\BibitemShut {NoStop}%
\bibitem [{\citenamefont {Proesmans}\ and\ \citenamefont {Van~den
  Broeck}(2017)}]{pro17}%
  \BibitemOpen
  \bibfield  {author} {\bibinfo {author} {\bibfnamefont {K.}~\bibnamefont
  {Proesmans}}\ and\ \bibinfo {author} {\bibfnamefont {C.}~\bibnamefont
  {Van~den Broeck}},\ }\href@noop {} {\bibfield  {journal} {\bibinfo  {journal}
  {EPL}\ }\textbf {\bibinfo {volume} {119}},\ \bibinfo {pages} {20001}
  (\bibinfo {year} {2017})}\BibitemShut {NoStop}%
\bibitem [{\citenamefont {Dechant}\ and\ \citenamefont {Sasa}(2018)}]{dec18}%
  \BibitemOpen
  \bibfield  {author} {\bibinfo {author} {\bibfnamefont {A.}~\bibnamefont
  {Dechant}}\ and\ \bibinfo {author} {\bibfnamefont {S.}~\bibnamefont {Sasa}},\
  }\href@noop {} {\bibfield  {journal} {\bibinfo  {journal} {Phys. Rev. E}\
  }\textbf {\bibinfo {volume} {97}},\ \bibinfo {pages} {062101} (\bibinfo
  {year} {2018})}\BibitemShut {NoStop}%
\bibitem [{\citenamefont {Falasco}\ \emph
  {et~al.}(2019{\natexlab{a}})\citenamefont {Falasco}, \citenamefont
  {Esposito},\ and\ \citenamefont {Delvenne}}]{fal19}%
  \BibitemOpen
  \bibfield  {author} {\bibinfo {author} {\bibfnamefont {G.}~\bibnamefont
  {Falasco}}, \bibinfo {author} {\bibfnamefont {M.}~\bibnamefont {Esposito}}, \
  and\ \bibinfo {author} {\bibfnamefont {J.-C.}\ \bibnamefont {Delvenne}},\
  }\href@noop {} {\bibfield  {journal} {\bibinfo  {journal} {arXiv:1906.11360}\
  } (\bibinfo {year} {2019}{\natexlab{a}})}\BibitemShut {NoStop}%
\bibitem [{\citenamefont {Van~Vu}\ and\ \citenamefont
  {Hasegawa}(2020)}]{van20}%
  \BibitemOpen
  \bibfield  {author} {\bibinfo {author} {\bibfnamefont {T.}~\bibnamefont
  {Van~Vu}}\ and\ \bibinfo {author} {\bibfnamefont {Y.}~\bibnamefont
  {Hasegawa}},\ }\href@noop {} {\bibfield  {journal} {\bibinfo  {journal}
  {Phys. Rev. Research}\ }\textbf {\bibinfo {volume} {2}},\ \bibinfo {pages}
  {013060} (\bibinfo {year} {2020})}\BibitemShut {NoStop}%
\bibitem [{\citenamefont {Falasco}\ \emph
  {et~al.}(2019{\natexlab{b}})\citenamefont {Falasco}, \citenamefont
  {Cossetto}, \citenamefont {Penocchio},\ and\ \citenamefont
  {Esposito}}]{fal19n}%
  \BibitemOpen
  \bibfield  {author} {\bibinfo {author} {\bibfnamefont {G.}~\bibnamefont
  {Falasco}}, \bibinfo {author} {\bibfnamefont {T.}~\bibnamefont {Cossetto}},
  \bibinfo {author} {\bibfnamefont {E.}~\bibnamefont {Penocchio}}, \ and\
  \bibinfo {author} {\bibfnamefont {M.}~\bibnamefont {Esposito}},\ }\href@noop
  {} {\bibfield  {journal} {\bibinfo  {journal} {New J. Phys.}\ }\textbf
  {\bibinfo {volume} {21}},\ \bibinfo {pages} {073005} (\bibinfo {year}
  {2019}{\natexlab{b}})}\BibitemShut {NoStop}%
\bibitem [{\citenamefont {Marsland~III}\ \emph {et~al.}(2019)\citenamefont
  {Marsland~III}, \citenamefont {Cui},\ and\ \citenamefont {Horowitz}}]{mar19}%
  \BibitemOpen
  \bibfield  {author} {\bibinfo {author} {\bibfnamefont {R.}~\bibnamefont
  {Marsland~III}}, \bibinfo {author} {\bibfnamefont {W.}~\bibnamefont {Cui}}, \
  and\ \bibinfo {author} {\bibfnamefont {J.~M.}\ \bibnamefont {Horowitz}},\
  }\href@noop {} {\bibfield  {journal} {\bibinfo  {journal} {J. R. Soc.
  Interface}\ }\textbf {\bibinfo {volume} {16}},\ \bibinfo {pages} {20190098}
  (\bibinfo {year} {2019})}\BibitemShut {NoStop}%
\bibitem [{\citenamefont {Horowitz}\ and\ \citenamefont
  {Gingrich}(2019)}]{hor19}%
  \BibitemOpen
  \bibfield  {author} {\bibinfo {author} {\bibfnamefont {J.~M.}\ \bibnamefont
  {Horowitz}}\ and\ \bibinfo {author} {\bibfnamefont {T.~R.}\ \bibnamefont
  {Gingrich}},\ }\href@noop {} {\bibfield  {journal} {\bibinfo  {journal} {Nat.
  Phys.}\ ,\ \bibinfo {pages} {1}} (\bibinfo {year} {2019})}\BibitemShut
  {NoStop}%
\bibitem [{\citenamefont {Shanahan}\ \emph {et~al.}(2018)\citenamefont
  {Shanahan}, \citenamefont {Chenu}, \citenamefont {Margolus},\ and\
  \citenamefont {Del~Campo}}]{sha18}%
  \BibitemOpen
  \bibfield  {author} {\bibinfo {author} {\bibfnamefont {B.}~\bibnamefont
  {Shanahan}}, \bibinfo {author} {\bibfnamefont {A.}~\bibnamefont {Chenu}},
  \bibinfo {author} {\bibfnamefont {N.}~\bibnamefont {Margolus}}, \ and\
  \bibinfo {author} {\bibfnamefont {A.}~\bibnamefont {Del~Campo}},\ }\href@noop
  {} {\bibfield  {journal} {\bibinfo  {journal} {Phys Rev. Lett.}\ }\textbf
  {\bibinfo {volume} {120}},\ \bibinfo {pages} {070401} (\bibinfo {year}
  {2018})}\BibitemShut {NoStop}%
\bibitem [{\citenamefont {Okuyama}\ and\ \citenamefont {Ohzeki}(2018)}]{oku18}%
  \BibitemOpen
  \bibfield  {author} {\bibinfo {author} {\bibfnamefont {M.}~\bibnamefont
  {Okuyama}}\ and\ \bibinfo {author} {\bibfnamefont {M.}~\bibnamefont
  {Ohzeki}},\ }\href@noop {} {\bibfield  {journal} {\bibinfo  {journal} {Phys
  Rev. Lett.}\ }\textbf {\bibinfo {volume} {120}},\ \bibinfo {pages} {070402}
  (\bibinfo {year} {2018})}\BibitemShut {NoStop}%
\bibitem [{\citenamefont {Deffner}\ and\ \citenamefont
  {Campbell}(2017)}]{def17}%
  \BibitemOpen
  \bibfield  {author} {\bibinfo {author} {\bibfnamefont {S.}~\bibnamefont
  {Deffner}}\ and\ \bibinfo {author} {\bibfnamefont {S.}~\bibnamefont
  {Campbell}},\ }\href@noop {} {\bibfield  {journal} {\bibinfo  {journal} {J.
  Phys. A}\ }\textbf {\bibinfo {volume} {50}},\ \bibinfo {pages} {453001}
  (\bibinfo {year} {2017})}\BibitemShut {NoStop}%
\bibitem [{\citenamefont {Shiraishi}\ \emph {et~al.}(2018)\citenamefont
  {Shiraishi}, \citenamefont {Funo},\ and\ \citenamefont {Saito}}]{shi18}%
  \BibitemOpen
  \bibfield  {author} {\bibinfo {author} {\bibfnamefont {N.}~\bibnamefont
  {Shiraishi}}, \bibinfo {author} {\bibfnamefont {K.}~\bibnamefont {Funo}}, \
  and\ \bibinfo {author} {\bibfnamefont {K.}~\bibnamefont {Saito}},\
  }\href@noop {} {\bibfield  {journal} {\bibinfo  {journal} {Phys. Rev. Lett.}\
  }\textbf {\bibinfo {volume} {121}},\ \bibinfo {pages} {070601} (\bibinfo
  {year} {2018})}\BibitemShut {NoStop}%
\bibitem [{\citenamefont {Ito}(2018)}]{ito18}%
  \BibitemOpen
  \bibfield  {author} {\bibinfo {author} {\bibfnamefont {S.}~\bibnamefont
  {Ito}},\ }\href@noop {} {\bibfield  {journal} {\bibinfo  {journal} {Phys.
  Rev. Lett.}\ }\textbf {\bibinfo {volume} {121}},\ \bibinfo {pages} {030605}
  (\bibinfo {year} {2018})}\BibitemShut {NoStop}%
\bibitem [{\citenamefont {Ito}\ and\ \citenamefont {Dechant}(2018)}]{ito18b}%
  \BibitemOpen
  \bibfield  {author} {\bibinfo {author} {\bibfnamefont {S.}~\bibnamefont
  {Ito}}\ and\ \bibinfo {author} {\bibfnamefont {A.}~\bibnamefont {Dechant}},\
  }\href@noop {} {\bibfield  {journal} {\bibinfo  {journal} {arXiv:1810.06832}\
  } (\bibinfo {year} {2018})}\BibitemShut {NoStop}%
\bibitem [{\citenamefont {Nicholson}\ \emph {et~al.}(2020)\citenamefont
  {Nicholson}, \citenamefont {Garcia-Pintos}, \citenamefont {del Campo},\ and\
  \citenamefont {Green}}]{nic20}%
  \BibitemOpen
  \bibfield  {author} {\bibinfo {author} {\bibfnamefont {S.~B.}\ \bibnamefont
  {Nicholson}}, \bibinfo {author} {\bibfnamefont {L.~P.}\ \bibnamefont
  {Garcia-Pintos}}, \bibinfo {author} {\bibfnamefont {A.}~\bibnamefont {del
  Campo}}, \ and\ \bibinfo {author} {\bibfnamefont {J.~R.}\ \bibnamefont
  {Green}},\ }\href@noop {} {\bibfield  {journal} {\bibinfo  {journal}
  {arXiv:2001.05418}\ } (\bibinfo {year} {2020})}\BibitemShut {NoStop}%
\bibitem [{\citenamefont {Gnesotto}\ \emph {et~al.}(2018)\citenamefont
  {Gnesotto}, \citenamefont {Mura}, \citenamefont {Gladrow},\ and\
  \citenamefont {Broedersz}}]{gne18}%
  \BibitemOpen
  \bibfield  {author} {\bibinfo {author} {\bibfnamefont {F.}~\bibnamefont
  {Gnesotto}}, \bibinfo {author} {\bibfnamefont {F.}~\bibnamefont {Mura}},
  \bibinfo {author} {\bibfnamefont {J.}~\bibnamefont {Gladrow}}, \ and\
  \bibinfo {author} {\bibfnamefont {C.}~\bibnamefont {Broedersz}},\ }\href@noop
  {} {\bibfield  {journal} {\bibinfo  {journal} {Rep. Prog. Phys.}\ }\textbf
  {\bibinfo {volume} {81}},\ \bibinfo {pages} {066601} (\bibinfo {year}
  {2018})}\BibitemShut {NoStop}%
\bibitem [{\citenamefont {Battle}\ \emph {et~al.}(2016)\citenamefont {Battle},
  \citenamefont {Broedersz}, \citenamefont {Fakhri}, \citenamefont {Geyer},
  \citenamefont {Howard}, \citenamefont {Schmidt},\ and\ \citenamefont
  {MacKintosh}}]{bat16}%
  \BibitemOpen
  \bibfield  {author} {\bibinfo {author} {\bibfnamefont {C.}~\bibnamefont
  {Battle}}, \bibinfo {author} {\bibfnamefont {C.~P.}\ \bibnamefont
  {Broedersz}}, \bibinfo {author} {\bibfnamefont {N.}~\bibnamefont {Fakhri}},
  \bibinfo {author} {\bibfnamefont {V.~F.}\ \bibnamefont {Geyer}}, \bibinfo
  {author} {\bibfnamefont {J.}~\bibnamefont {Howard}}, \bibinfo {author}
  {\bibfnamefont {C.~F.}\ \bibnamefont {Schmidt}}, \ and\ \bibinfo {author}
  {\bibfnamefont {F.~C.}\ \bibnamefont {MacKintosh}},\ }\href@noop {}
  {\bibfield  {journal} {\bibinfo  {journal} {Science}\ }\textbf {\bibinfo
  {volume} {352}},\ \bibinfo {pages} {604} (\bibinfo {year}
  {2016})}\BibitemShut {NoStop}%
\bibitem [{\citenamefont {Li}\ \emph {et~al.}(2019)\citenamefont {Li},
  \citenamefont {Horowitz}, \citenamefont {Gingrich},\ and\ \citenamefont
  {Fakhri}}]{li19}%
  \BibitemOpen
  \bibfield  {author} {\bibinfo {author} {\bibfnamefont {J.}~\bibnamefont
  {Li}}, \bibinfo {author} {\bibfnamefont {J.~M.}\ \bibnamefont {Horowitz}},
  \bibinfo {author} {\bibfnamefont {T.~R.}\ \bibnamefont {Gingrich}}, \ and\
  \bibinfo {author} {\bibfnamefont {N.}~\bibnamefont {Fakhri}},\ }\href@noop {}
  {\bibfield  {journal} {\bibinfo  {journal} {Nat. Com.}\ }\textbf {\bibinfo
  {volume} {10}},\ \bibinfo {pages} {1} (\bibinfo {year} {2019})}\BibitemShut
  {NoStop}%
\bibitem [{\citenamefont {Gingrich}\ and\ \citenamefont
  {Horowitz}(2017)}]{gin17fp}%
  \BibitemOpen
  \bibfield  {author} {\bibinfo {author} {\bibfnamefont {T.~R.}\ \bibnamefont
  {Gingrich}}\ and\ \bibinfo {author} {\bibfnamefont {J.~M.}\ \bibnamefont
  {Horowitz}},\ }\href@noop {} {\bibfield  {journal} {\bibinfo  {journal}
  {Phys. Rev. Lett.}\ }\textbf {\bibinfo {volume} {119}},\ \bibinfo {pages}
  {170601} (\bibinfo {year} {2017})}\BibitemShut {NoStop}%
\bibitem [{\citenamefont {Garrahan}(2017)}]{gar17}%
  \BibitemOpen
  \bibfield  {author} {\bibinfo {author} {\bibfnamefont {J.~P.}\ \bibnamefont
  {Garrahan}},\ }\href@noop {} {\bibfield  {journal} {\bibinfo  {journal}
  {Phys. Rev. E}\ }\textbf {\bibinfo {volume} {95}},\ \bibinfo {pages} {032134}
  (\bibinfo {year} {2017})}\BibitemShut {NoStop}%
\bibitem [{\citenamefont {Neri}\ \emph {et~al.}(2017)\citenamefont {Neri},
  \citenamefont {Rold{\'a}n},\ and\ \citenamefont {J{\"u}licher}}]{ner17}%
  \BibitemOpen
  \bibfield  {author} {\bibinfo {author} {\bibfnamefont {I.}~\bibnamefont
  {Neri}}, \bibinfo {author} {\bibfnamefont {{\'E}.}~\bibnamefont
  {Rold{\'a}n}}, \ and\ \bibinfo {author} {\bibfnamefont {F.}~\bibnamefont
  {J{\"u}licher}},\ }\href@noop {} {\bibfield  {journal} {\bibinfo  {journal}
  {Phys. Rev. X}\ }\textbf {\bibinfo {volume} {7}},\ \bibinfo {pages} {011019}
  (\bibinfo {year} {2017})}\BibitemShut {NoStop}%
\bibitem [{\citenamefont {Neri}(2020)}]{ner20}%
  \BibitemOpen
  \bibfield  {author} {\bibinfo {author} {\bibfnamefont {I.}~\bibnamefont
  {Neri}},\ }\href@noop {} {\bibfield  {journal} {\bibinfo  {journal} {Phys.
  Rev. Lett.}\ }\textbf {\bibinfo {volume} {124}},\ \bibinfo {pages} {040601}
  (\bibinfo {year} {2020})}\BibitemShut {NoStop}%
\bibitem [{\citenamefont {Redner}(2001)}]{red01}%
  \BibitemOpen
  \bibfield  {author} {\bibinfo {author} {\bibfnamefont {S.}~\bibnamefont
  {Redner}},\ }\href@noop {} {\emph {\bibinfo {title} {A guide to first-passage
  processes}}}\ (\bibinfo  {publisher} {Cambridge University Press},\ \bibinfo
  {year} {2001})\BibitemShut {NoStop}%
\bibitem [{\citenamefont {Graham}(1987)}]{gra87}%
  \BibitemOpen
  \bibfield  {author} {\bibinfo {author} {\bibfnamefont {R.}~\bibnamefont
  {Graham}},\ }in\ \href@noop {} {\emph {\bibinfo {booktitle} {Fluctuations and
  Stochastic Phenomena in Condensed Matter}}}\ (\bibinfo  {publisher}
  {Springer},\ \bibinfo {year} {1987})\ pp.\ \bibinfo {pages}
  {1--34}\BibitemShut {NoStop}%
\bibitem [{\citenamefont {H{\"a}nggi}\ and\ \citenamefont
  {Jung}(1995)}]{han95}%
  \BibitemOpen
  \bibfield  {author} {\bibinfo {author} {\bibfnamefont {P.}~\bibnamefont
  {H{\"a}nggi}}\ and\ \bibinfo {author} {\bibfnamefont {P.}~\bibnamefont
  {Jung}},\ }\href@noop {} {\bibfield  {journal} {\bibinfo  {journal} {Adv.
  Chem. Phys}\ }\textbf {\bibinfo {volume} {89}},\ \bibinfo {pages} {239}
  (\bibinfo {year} {1995})}\BibitemShut {NoStop}%
\bibitem [{\citenamefont {Bertini}\ \emph {et~al.}(2015)\citenamefont
  {Bertini}, \citenamefont {De~Sole}, \citenamefont {Gabrielli}, \citenamefont
  {Jona-Lasinio},\ and\ \citenamefont {Landim}}]{ber15}%
  \BibitemOpen
  \bibfield  {author} {\bibinfo {author} {\bibfnamefont {L.}~\bibnamefont
  {Bertini}}, \bibinfo {author} {\bibfnamefont {A.}~\bibnamefont {De~Sole}},
  \bibinfo {author} {\bibfnamefont {D.}~\bibnamefont {Gabrielli}}, \bibinfo
  {author} {\bibfnamefont {G.}~\bibnamefont {Jona-Lasinio}}, \ and\ \bibinfo
  {author} {\bibfnamefont {C.}~\bibnamefont {Landim}},\ }\href@noop {}
  {\bibfield  {journal} {\bibinfo  {journal} {Rev. Mod. Phys.}\ }\textbf
  {\bibinfo {volume} {87}},\ \bibinfo {pages} {593} (\bibinfo {year}
  {2015})}\BibitemShut {NoStop}%
\bibitem [{\citenamefont {Speck}\ and\ \citenamefont {Seifert}(2007)}]{spe07}%
  \BibitemOpen
  \bibfield  {author} {\bibinfo {author} {\bibfnamefont {T.}~\bibnamefont
  {Speck}}\ and\ \bibinfo {author} {\bibfnamefont {U.}~\bibnamefont
  {Seifert}},\ }\href@noop {} {\bibfield  {journal} {\bibinfo  {journal} {J.
  Stat. Mech.}\ }\textbf {\bibinfo {volume} {2007}},\ \bibinfo {pages} {L09002}
  (\bibinfo {year} {2007})}\BibitemShut {NoStop}%
\bibitem [{\citenamefont {Esposito}\ and\ \citenamefont
  {Lindenberg}(2008)}]{esp08}%
  \BibitemOpen
  \bibfield  {author} {\bibinfo {author} {\bibfnamefont {M.}~\bibnamefont
  {Esposito}}\ and\ \bibinfo {author} {\bibfnamefont {K.}~\bibnamefont
  {Lindenberg}},\ }\href@noop {} {\bibfield  {journal} {\bibinfo  {journal}
  {Phys. Rev. E}\ }\textbf {\bibinfo {volume} {77}},\ \bibinfo {pages} {051119}
  (\bibinfo {year} {2008})}\BibitemShut {NoStop}%
\bibitem [{\citenamefont {Carollo}\ \emph {et~al.}(2019)\citenamefont
  {Carollo}, \citenamefont {Jack},\ and\ \citenamefont {Garrahan}}]{car19}%
  \BibitemOpen
  \bibfield  {author} {\bibinfo {author} {\bibfnamefont {F.}~\bibnamefont
  {Carollo}}, \bibinfo {author} {\bibfnamefont {R.~L.}\ \bibnamefont {Jack}}, \
  and\ \bibinfo {author} {\bibfnamefont {J.~P.}\ \bibnamefont {Garrahan}},\
  }\href@noop {} {\bibfield  {journal} {\bibinfo  {journal} {Phys. Rev. Lett.}\
  }\textbf {\bibinfo {volume} {122}},\ \bibinfo {pages} {130605} (\bibinfo
  {year} {2019})}\BibitemShut {NoStop}%
\bibitem [{\citenamefont {Seifert}(2012)}]{sei12}%
  \BibitemOpen
  \bibfield  {author} {\bibinfo {author} {\bibfnamefont {U.}~\bibnamefont
  {Seifert}},\ }\href@noop {} {\bibfield  {journal} {\bibinfo  {journal} {Rep.
  Prog. Phys.}\ }\textbf {\bibinfo {volume} {75}},\ \bibinfo {pages} {126001}
  (\bibinfo {year} {2012})}\BibitemShut {NoStop}%
\bibitem [{\citenamefont {Van~den Broeck}\ and\ \citenamefont
  {Esposito}(2015)}]{van15}%
  \BibitemOpen
  \bibfield  {author} {\bibinfo {author} {\bibfnamefont {C.}~\bibnamefont
  {Van~den Broeck}}\ and\ \bibinfo {author} {\bibfnamefont {M.}~\bibnamefont
  {Esposito}},\ }\href@noop {} {\bibfield  {journal} {\bibinfo  {journal}
  {Physica A}\ }\textbf {\bibinfo {volume} {418}},\ \bibinfo {pages} {6}
  (\bibinfo {year} {2015})}\BibitemShut {NoStop}%
\bibitem [{\citenamefont {Murashita}\ \emph {et~al.}(2014)\citenamefont
  {Murashita}, \citenamefont {Funo},\ and\ \citenamefont {Ueda}}]{mur14}%
  \BibitemOpen
  \bibfield  {author} {\bibinfo {author} {\bibfnamefont {Y.}~\bibnamefont
  {Murashita}}, \bibinfo {author} {\bibfnamefont {K.}~\bibnamefont {Funo}}, \
  and\ \bibinfo {author} {\bibfnamefont {M.}~\bibnamefont {Ueda}},\ }\href@noop
  {} {\bibfield  {journal} {\bibinfo  {journal} {Phys. Rev. E}\ }\textbf
  {\bibinfo {volume} {90}},\ \bibinfo {pages} {042110} (\bibinfo {year}
  {2014})}\BibitemShut {NoStop}%
\bibitem [{\citenamefont {Funo}\ \emph {et~al.}(2015)\citenamefont {Funo},
  \citenamefont {Murashita},\ and\ \citenamefont {Ueda}}]{fun15}%
  \BibitemOpen
  \bibfield  {author} {\bibinfo {author} {\bibfnamefont {K.}~\bibnamefont
  {Funo}}, \bibinfo {author} {\bibfnamefont {Y.}~\bibnamefont {Murashita}}, \
  and\ \bibinfo {author} {\bibfnamefont {M.}~\bibnamefont {Ueda}},\ }\href@noop
  {} {\bibfield  {journal} {\bibinfo  {journal} {New J. Phys.}\ }\textbf
  {\bibinfo {volume} {17}},\ \bibinfo {pages} {075005} (\bibinfo {year}
  {2015})}\BibitemShut {NoStop}%
\bibitem [{\citenamefont {Murashita}\ \emph {et~al.}(2018)\citenamefont
  {Murashita}, \citenamefont {Kura},\ and\ \citenamefont {Ueda}}]{mur18}%
  \BibitemOpen
  \bibfield  {author} {\bibinfo {author} {\bibfnamefont {Y.}~\bibnamefont
  {Murashita}}, \bibinfo {author} {\bibfnamefont {N.}~\bibnamefont {Kura}}, \
  and\ \bibinfo {author} {\bibfnamefont {M.}~\bibnamefont {Ueda}},\ }\href@noop
  {} {\bibfield  {journal} {\bibinfo  {journal} {arXiv:1802.10483}\ } (\bibinfo
  {year} {2018})}\BibitemShut {NoStop}%
\bibitem [{sup()}]{supinf}%
  \BibitemOpen
  \href@noop {} {}\bibinfo {note} {See Supplemental Material for the details of
  the mathematical derivations}\BibitemShut {NoStop}%
\bibitem [{\citenamefont {Forbes}\ \emph {et~al.}(2011)\citenamefont {Forbes},
  \citenamefont {Evans}, \citenamefont {Hastings},\ and\ \citenamefont
  {Peacock}}]{for11}%
  \BibitemOpen
  \bibfield  {author} {\bibinfo {author} {\bibfnamefont {C.}~\bibnamefont
  {Forbes}}, \bibinfo {author} {\bibfnamefont {M.}~\bibnamefont {Evans}},
  \bibinfo {author} {\bibfnamefont {N.}~\bibnamefont {Hastings}}, \ and\
  \bibinfo {author} {\bibfnamefont {B.}~\bibnamefont {Peacock}},\ }\href@noop
  {} {\emph {\bibinfo {title} {Statistical distributions}}}\ (\bibinfo
  {publisher} {John Wiley \& Sons},\ \bibinfo {year} {2011})\BibitemShut
  {NoStop}%
\bibitem [{\citenamefont {J{\"u}licher}\ \emph {et~al.}(1997)\citenamefont
  {J{\"u}licher}, \citenamefont {Ajdari},\ and\ \citenamefont {Prost}}]{jul97}%
  \BibitemOpen
  \bibfield  {author} {\bibinfo {author} {\bibfnamefont {F.}~\bibnamefont
  {J{\"u}licher}}, \bibinfo {author} {\bibfnamefont {A.}~\bibnamefont
  {Ajdari}}, \ and\ \bibinfo {author} {\bibfnamefont {J.}~\bibnamefont
  {Prost}},\ }\href@noop {} {\bibfield  {journal} {\bibinfo  {journal} {Rev.
  Mod. Phys.}\ }\textbf {\bibinfo {volume} {69}},\ \bibinfo {pages} {1269}
  (\bibinfo {year} {1997})}\BibitemShut {NoStop}%
\bibitem [{\citenamefont {Dieterich}\ \emph {et~al.}(1980)\citenamefont
  {Dieterich}, \citenamefont {Fulde},\ and\ \citenamefont {Peschel}}]{die80}%
  \BibitemOpen
  \bibfield  {author} {\bibinfo {author} {\bibfnamefont {W.}~\bibnamefont
  {Dieterich}}, \bibinfo {author} {\bibfnamefont {P.}~\bibnamefont {Fulde}}, \
  and\ \bibinfo {author} {\bibfnamefont {I.}~\bibnamefont {Peschel}},\
  }\href@noop {} {\bibfield  {journal} {\bibinfo  {journal} {Adv. Phys.}\
  }\textbf {\bibinfo {volume} {29}},\ \bibinfo {pages} {527} (\bibinfo {year}
  {1980})}\BibitemShut {NoStop}%
\bibitem [{\citenamefont {Bouchet}\ and\ \citenamefont
  {Reygner}(2016)}]{bou16}%
  \BibitemOpen
  \bibfield  {author} {\bibinfo {author} {\bibfnamefont {F.}~\bibnamefont
  {Bouchet}}\ and\ \bibinfo {author} {\bibfnamefont {J.}~\bibnamefont
  {Reygner}},\ }in\ \href@noop {} {\emph {\bibinfo {booktitle} {Ann. Henri
  Poincar{\'e}}}},\ Vol.~\bibinfo {volume} {17}\ (\bibinfo {organization}
  {Springer},\ \bibinfo {year} {2016})\ pp.\ \bibinfo {pages}
  {3499--3532}\BibitemShut {NoStop}%
\bibitem [{\citenamefont {Risken}(1989)}]{ris89}%
  \BibitemOpen
  \bibfield  {author} {\bibinfo {author} {\bibfnamefont {H.}~\bibnamefont
  {Risken}},\ }\href@noop {} {\emph {\bibinfo {title} {The {F}okker-{P}lanck
  Equation}}},\ \bibinfo {edition} {2nd}\ ed.\ (\bibinfo  {publisher}
  {Springer-Verlag},\ \bibinfo {address} {Berlin},\ \bibinfo {year}
  {1989})\BibitemShut {NoStop}%
\bibitem [{\citenamefont {H{\"a}nggi}\ \emph {et~al.}(1990)\citenamefont
  {H{\"a}nggi}, \citenamefont {Talkner},\ and\ \citenamefont
  {Borkovec}}]{han90}%
  \BibitemOpen
  \bibfield  {author} {\bibinfo {author} {\bibfnamefont {P.}~\bibnamefont
  {H{\"a}nggi}}, \bibinfo {author} {\bibfnamefont {P.}~\bibnamefont {Talkner}},
  \ and\ \bibinfo {author} {\bibfnamefont {M.}~\bibnamefont {Borkovec}},\
  }\href@noop {} {\bibfield  {journal} {\bibinfo  {journal} {Rev. Mod. Phys.}\
  }\textbf {\bibinfo {volume} {62}},\ \bibinfo {pages} {251} (\bibinfo {year}
  {1990})}\BibitemShut {NoStop}%
\bibitem [{\citenamefont {Rao}\ and\ \citenamefont {Esposito}(2018)}]{rao18}%
  \BibitemOpen
  \bibfield  {author} {\bibinfo {author} {\bibfnamefont {R.}~\bibnamefont
  {Rao}}\ and\ \bibinfo {author} {\bibfnamefont {M.}~\bibnamefont {Esposito}},\
  }\href@noop {} {\bibfield  {journal} {\bibinfo  {journal} {Entropy}\ }\textbf
  {\bibinfo {volume} {20}},\ \bibinfo {pages} {635} (\bibinfo {year}
  {2018})}\BibitemShut {NoStop}%
\bibitem [{\citenamefont {Di~Terlizzi}\ and\ \citenamefont
  {Baiesi}(2018)}]{dit18}%
  \BibitemOpen
  \bibfield  {author} {\bibinfo {author} {\bibfnamefont {I.}~\bibnamefont
  {Di~Terlizzi}}\ and\ \bibinfo {author} {\bibfnamefont {M.}~\bibnamefont
  {Baiesi}},\ }\href@noop {} {\bibfield  {journal} {\bibinfo  {journal} {J.
  Phys. A: Math. Gen}\ }\textbf {\bibinfo {volume} {52}},\ \bibinfo {pages}
  {02LT03} (\bibinfo {year} {2018})}\BibitemShut {NoStop}%
\bibitem [{\citenamefont {Breymann}\ \emph {et~al.}(1996)\citenamefont
  {Breymann}, \citenamefont {T{\'e}l},\ and\ \citenamefont {Vollmer}}]{bre96}%
  \BibitemOpen
  \bibfield  {author} {\bibinfo {author} {\bibfnamefont {W.}~\bibnamefont
  {Breymann}}, \bibinfo {author} {\bibfnamefont {T.}~\bibnamefont {T{\'e}l}}, \
  and\ \bibinfo {author} {\bibfnamefont {J.}~\bibnamefont {Vollmer}},\
  }\href@noop {} {\bibfield  {journal} {\bibinfo  {journal} {Phys. Rev. Lett.}\
  }\textbf {\bibinfo {volume} {77}},\ \bibinfo {pages} {2945} (\bibinfo {year}
  {1996})}\BibitemShut {NoStop}%
\bibitem [{\citenamefont {Gaspard}(2007)}]{gas07}%
  \BibitemOpen
  \bibfield  {author} {\bibinfo {author} {\bibfnamefont {P.}~\bibnamefont
  {Gaspard}},\ }\href@noop {} {\bibfield  {journal} {\bibinfo  {journal} {Adv.
  Chem. Phys.}\ }\textbf {\bibinfo {volume} {135}},\ \bibinfo {pages} {83}
  (\bibinfo {year} {2007})}\BibitemShut {NoStop}%
\bibitem [{\citenamefont {Gaspard}(2005)}]{gas05}%
  \BibitemOpen
  \bibfield  {author} {\bibinfo {author} {\bibfnamefont {P.}~\bibnamefont
  {Gaspard}},\ }\href@noop {} {\emph {\bibinfo {title} {Chaos, scattering and
  statistical mechanics}}},\ Vol.~\bibinfo {volume} {9}\ (\bibinfo  {publisher}
  {Cambridge University Press},\ \bibinfo {year} {2005})\BibitemShut {NoStop}%
\bibitem [{\citenamefont {{\v{S}}iler}\ \emph {et~al.}(2018)\citenamefont
  {{\v{S}}iler}, \citenamefont {Ornigotti}, \citenamefont {Brzobohat{\`y}},
  \citenamefont {J{\'a}kl}, \citenamefont {Ryabov}, \citenamefont {Holubec},
  \citenamefont {Zem{\'a}nek},\ and\ \citenamefont {Filip}}]{vsi18}%
  \BibitemOpen
  \bibfield  {author} {\bibinfo {author} {\bibfnamefont {M.}~\bibnamefont
  {{\v{S}}iler}}, \bibinfo {author} {\bibfnamefont {L.}~\bibnamefont
  {Ornigotti}}, \bibinfo {author} {\bibfnamefont {O.}~\bibnamefont
  {Brzobohat{\`y}}}, \bibinfo {author} {\bibfnamefont {P.}~\bibnamefont
  {J{\'a}kl}}, \bibinfo {author} {\bibfnamefont {A.}~\bibnamefont {Ryabov}},
  \bibinfo {author} {\bibfnamefont {V.}~\bibnamefont {Holubec}}, \bibinfo
  {author} {\bibfnamefont {P.}~\bibnamefont {Zem{\'a}nek}}, \ and\ \bibinfo
  {author} {\bibfnamefont {R.}~\bibnamefont {Filip}},\ }\href@noop {}
  {\bibfield  {journal} {\bibinfo  {journal} {Physical review letters}\
  }\textbf {\bibinfo {volume} {121}},\ \bibinfo {pages} {230601} (\bibinfo
  {year} {2018})}\BibitemShut {NoStop}%
\bibitem [{\citenamefont {Ornigotti}\ \emph {et~al.}(2018)\citenamefont
  {Ornigotti}, \citenamefont {Ryabov}, \citenamefont {Holubec},\ and\
  \citenamefont {Filip}}]{orn18}%
  \BibitemOpen
  \bibfield  {author} {\bibinfo {author} {\bibfnamefont {L.}~\bibnamefont
  {Ornigotti}}, \bibinfo {author} {\bibfnamefont {A.}~\bibnamefont {Ryabov}},
  \bibinfo {author} {\bibfnamefont {V.}~\bibnamefont {Holubec}}, \ and\
  \bibinfo {author} {\bibfnamefont {R.}~\bibnamefont {Filip}},\ }\href@noop {}
  {\bibfield  {journal} {\bibinfo  {journal} {Physical Review E}\ }\textbf
  {\bibinfo {volume} {97}},\ \bibinfo {pages} {032127} (\bibinfo {year}
  {2018})}\BibitemShut {NoStop}%
\bibitem [{\citenamefont {Mehta}\ and\ \citenamefont {Schwab}(2012)}]{meh12}%
  \BibitemOpen
  \bibfield  {author} {\bibinfo {author} {\bibfnamefont {P.}~\bibnamefont
  {Mehta}}\ and\ \bibinfo {author} {\bibfnamefont {D.~J.}\ \bibnamefont
  {Schwab}},\ }\href@noop {} {\bibfield  {journal} {\bibinfo  {journal} {Proc.
  Natl. Acad. Sci.}\ }\textbf {\bibinfo {volume} {109}},\ \bibinfo {pages}
  {17978} (\bibinfo {year} {2012})}\BibitemShut {NoStop}%
\bibitem [{\citenamefont {Diana}\ \emph {et~al.}(2013)\citenamefont {Diana},
  \citenamefont {Bagci},\ and\ \citenamefont {Esposito}}]{dia13}%
  \BibitemOpen
  \bibfield  {author} {\bibinfo {author} {\bibfnamefont {G.}~\bibnamefont
  {Diana}}, \bibinfo {author} {\bibfnamefont {G.~B.}\ \bibnamefont {Bagci}}, \
  and\ \bibinfo {author} {\bibfnamefont {M.}~\bibnamefont {Esposito}},\
  }\href@noop {} {\bibfield  {journal} {\bibinfo  {journal} {Phys. Rev. E}\
  }\textbf {\bibinfo {volume} {87}},\ \bibinfo {pages} {012111} (\bibinfo
  {year} {2013})}\BibitemShut {NoStop}%
\bibitem [{\citenamefont {Sartori}\ \emph {et~al.}(2014)\citenamefont
  {Sartori}, \citenamefont {Granger}, \citenamefont {Lee},\ and\ \citenamefont
  {Horowitz}}]{sar14}%
  \BibitemOpen
  \bibfield  {author} {\bibinfo {author} {\bibfnamefont {P.}~\bibnamefont
  {Sartori}}, \bibinfo {author} {\bibfnamefont {L.}~\bibnamefont {Granger}},
  \bibinfo {author} {\bibfnamefont {C.~F.}\ \bibnamefont {Lee}}, \ and\
  \bibinfo {author} {\bibfnamefont {J.~M.}\ \bibnamefont {Horowitz}},\
  }\href@noop {} {\bibfield  {journal} {\bibinfo  {journal} {PLOS Comput.
  Biol.}\ }\textbf {\bibinfo {volume} {10}} (\bibinfo {year}
  {2014})}\BibitemShut {NoStop}%
\bibitem [{\citenamefont {Lan}\ and\ \citenamefont {Tu}(2016)}]{ian16}%
  \BibitemOpen
  \bibfield  {author} {\bibinfo {author} {\bibfnamefont {G.}~\bibnamefont
  {Lan}}\ and\ \bibinfo {author} {\bibfnamefont {Y.}~\bibnamefont {Tu}},\
  }\href@noop {} {\bibfield  {journal} {\bibinfo  {journal} {Rep. Prog. Phys.}\
  }\textbf {\bibinfo {volume} {79}},\ \bibinfo {pages} {052601} (\bibinfo
  {year} {2016})}\BibitemShut {NoStop}%
\bibitem [{\citenamefont {Horowitz}\ and\ \citenamefont
  {Esposito}(2014)}]{hor14}%
  \BibitemOpen
  \bibfield  {author} {\bibinfo {author} {\bibfnamefont {J.~M.}\ \bibnamefont
  {Horowitz}}\ and\ \bibinfo {author} {\bibfnamefont {M.}~\bibnamefont
  {Esposito}},\ }\href@noop {} {\bibfield  {journal} {\bibinfo  {journal}
  {Phys. Rev. X}\ }\textbf {\bibinfo {volume} {4}},\ \bibinfo {pages} {031015}
  (\bibinfo {year} {2014})}\BibitemShut {NoStop}%
\bibitem [{\citenamefont {Parrondo}\ \emph {et~al.}(2015)\citenamefont
  {Parrondo}, \citenamefont {Horowitz},\ and\ \citenamefont {Sagawa}}]{par15}%
  \BibitemOpen
  \bibfield  {author} {\bibinfo {author} {\bibfnamefont {J.~M.}\ \bibnamefont
  {Parrondo}}, \bibinfo {author} {\bibfnamefont {J.~M.}\ \bibnamefont
  {Horowitz}}, \ and\ \bibinfo {author} {\bibfnamefont {T.}~\bibnamefont
  {Sagawa}},\ }\href@noop {} {\bibfield  {journal} {\bibinfo  {journal} {Nat.
  Phys.}\ }\textbf {\bibinfo {volume} {11}},\ \bibinfo {pages} {131} (\bibinfo
  {year} {2015})}\BibitemShut {NoStop}%
\bibitem [{\citenamefont {Kolchinsky}\ and\ \citenamefont
  {Wolpert}(2018)}]{kol18}%
  \BibitemOpen
  \bibfield  {author} {\bibinfo {author} {\bibfnamefont {A.}~\bibnamefont
  {Kolchinsky}}\ and\ \bibinfo {author} {\bibfnamefont {D.~H.}\ \bibnamefont
  {Wolpert}},\ }\href@noop {} {\bibfield  {journal} {\bibinfo  {journal}
  {Interface Focus}\ }\textbf {\bibinfo {volume} {8}},\ \bibinfo {pages}
  {20180041} (\bibinfo {year} {2018})}\BibitemShut {NoStop}%
\bibitem [{\citenamefont {Deffner}(2020)}]{deff20}%
  \BibitemOpen
  \bibfield  {author} {\bibinfo {author} {\bibfnamefont {S.}~\bibnamefont
  {Deffner}},\ }\href@noop {} {\bibfield  {journal} {\bibinfo  {journal} {Phys.
  Rev. Research}\ }\textbf {\bibinfo {volume} {2}},\ \bibinfo {pages} {013161}
  (\bibinfo {year} {2020})}\BibitemShut {NoStop}%
\bibitem [{\citenamefont {T{\'e}l}\ \emph {et~al.}(1996)\citenamefont
  {T{\'e}l}, \citenamefont {Vollmer},\ and\ \citenamefont {Breymann}}]{tel96}%
  \BibitemOpen
  \bibfield  {author} {\bibinfo {author} {\bibfnamefont {T.}~\bibnamefont
  {T{\'e}l}}, \bibinfo {author} {\bibfnamefont {J.}~\bibnamefont {Vollmer}}, \
  and\ \bibinfo {author} {\bibfnamefont {W.}~\bibnamefont {Breymann}},\
  }\href@noop {} {\bibfield  {journal} {\bibinfo  {journal} {EPL}\ }\textbf
  {\bibinfo {volume} {35}},\ \bibinfo {pages} {659} (\bibinfo {year}
  {1996})}\BibitemShut {NoStop}%
\end{thebibliography}%

\begin{widetext}

\newpage 
\appendix 
 \section{Derivation of the bound}

We consider stochastic trajectories $\omega_t= \{ x_{t'} :  0 \ge t' \ge t \}$  of duration $t$ corresponding to an ordered list of states $x_{t'}$. On the trajectory space  $\Omega_t \ni \omega_t$ we consider the probability measure $P(\omega_t)= P(\omega_t|x_0) \rho(x_0)$, with $\rho(x_0)$ the stationary probability density of state $x_0$. 
The stopping time $\tau:= \text{inf} \{t   \geq 0 : O(\omega_t) \in D\}$ is the minimum time for an observable $O: \Omega_t \to \mathbb{R}$ to reach values belonging to a specific domain $D \subset \mathbb{R}$, loosely referred to as `threshold'. The observable $O$, the threshold $D$, and the time $\tau$, define the physical process and its duration.

We identify the space of `survived' trajectories at time $t$, $\Omega^{\text{s}}_t$, such that if $\omega_t \in \Omega^{\text{s}}_t$ then $O(\omega_t) \not \in D$. They correspond to trajectories in which the process is not completed. The associated probability that the process is not yet completed at time $t$ is expressed by the survival probability $p^\text{s}(t):= \text{Prob}(\tau > t) $, which by definition satisfies $p^\text{s}(0)=1$. We can formally write it as 
\begin{align}\label{ps}
p^\text{s}(t) =  \sum_{\omega_t \in \Omega_t^{\text{s}}} P(\omega_t)=  \sum_{\omega_t \in \Omega_t } \chi(O(\omega_t)) ,P(\omega_t).
\end{align}
where $ \chi(O(\omega_t))$ is an indicator function giving 1 if the process defined by $O$ is not complete and 0 otherwise:
\begin{align}
\chi(O)= 
\begin{cases}
1 & \text{if } O \not\in D \\
0 & \text{if } O  \in D
\end{cases}
\end{align}
We then reweight the average in \eqref{ps} considering the time reversed trajectories $\tilde \omega_t =\{ \theta x_{t-t'}: 0 \ge t' \ge t\} $, where $\theta$ accounts for the parity of $x$:
\begin{align}\label{ps2}
p^{\text{s}}(t)= \sum_{\omega_t \in \Omega_t } \chi(O(\omega_t)) \frac{P(\omega_t)}{P(\tilde \omega_t)} P(\tilde \omega_t) .
\end{align}
Here $P(\tilde \omega_t) = P(\tilde \omega_t| \tilde x_0) \rho(\tilde x_0) $ is the probability of time reversed trajectories starting from the  stationary probability $\rho(\tilde x_0)$.
When the dynamics satisfies local detailed balance, we can introduce the entropy production rate $\dot \Sigma(\tilde \omega_t)$ of the stationary time-reversed dynamics as
\begin{align}
\frac{P(\omega_t)}{P(\tilde \omega_t)} &= e^{-\log \frac{P(\tilde \omega_t| \tilde x_0)}{P(\omega_t| x_0)} + \log \rho(\tilde x_t) -\log \rho(\tilde x_0) }\\
&= e^{-(S_e(\tilde \omega_t) + \Delta S(\tilde \omega_t))/k_\text{B} }\\
&= e^{-\frac{1}{k_\text{B}} \int_0^t (\dot S_e +\frac{d S}{dt'})(\tilde \omega_t) }\\
&= e^{-\frac{1}{k_\text{B}} \int_0^t \dot \Sigma (\tilde \omega_t) }. \label{epr}
\end{align}
Here we have used the decomposition of $\dot \Sigma$ at time $t'$ into the entropy flow in the reservoirs, $\dot S_\text{e}(\tilde \omega_t):=  k_\text{B} \log \frac{P(\tilde \omega_t| \tilde x_0)}{P(\omega_t| x_0)} $, and the derivative of the system entropy $S := -\log\rho(x_{t'})$.

Using \eqref{epr} and rewriting all terms appearing in \eqref{ps2} in terms of reversed trajectories we find
\begin{align}
p^{\text{s}}(t)&= \sum_{\omega_t \in \Omega_t } \chi(O( \omega_t))e^{-\frac{1}{k_\text{B}} \int_0^t \dot \Sigma (\tilde \omega_t) } P(\tilde \omega_t) \\
&= \sum_{\tilde \omega_t \in \Omega_t } \chi(O( \omega_t))e^{-\frac{1}{k_\text{B}} \int_0^t \dot \Sigma (\tilde \omega_t) } P(\tilde \omega_t) \\
&= \sum_{\omega_t \in \Omega_t } \chi(\tilde O( \omega_t))e^{-\frac{1}{k_\text{B}} \int_0^t \dot \Sigma ( \omega_t) } P( \omega_t) \label{meantilde} .
\end{align}
For the second equality we used that $\sum_{\omega_t} = \sum_{\tilde {\omega_t}} $ and for the third we renamed $\tilde \omega_t$ as $\omega_t$ and defined the time reversed observable $\tilde O(\omega_t): =O(\tilde  \omega_t)$. For example, if $O$ is an integrated current  in $[0,t]$, then $\tilde O(\omega_t)= - O( \omega_t)$ and the set of time-reversed survived trajectories $\tilde \Omega_t^\text{s}$ is different from $\Omega_t^\text{s}$. If, instead, the observable is time-symmetric, $O(\tilde \omega_t)= O( \omega_t)$, we have $\tilde \Omega_t^\text{s}= \Omega_t^\text{s}$.

In order to use Jensen's inequality, we define the normalization of the average in \eqref{meantilde},
\begin{align}\label{pstilde}
\tilde p^{\text{s}}(t)= \sum_{\omega_t \in \Omega_t } \chi(\tilde O( \omega_t)) P( \omega_t) = \sum_{\omega_t \in \tilde \Omega_t^\text{s}} P( \omega_t) ,
\end{align}
which is the survival probability of the process defined by $\tilde O$. Then we can normalize the average in \eqref{meantilde} multiplying and dividing by \eqref{pstilde},
\begin{align}\label{psps}
p^{\text{s}}(t)&=\tilde p^{\text{s}}(t)  \frac{\sum_{\omega_t \in \Omega_t } \chi(\tilde O(\omega_t))e^{-\frac{1}{k_\text{B}} \int_0^t \dot \Sigma ( \omega_t) }  P( \omega_t) }{\sum_{\omega_t \in \Omega_t  } \chi(\tilde O( \omega_t)) P( \omega_t)}.
\end{align}
Defining the normalized average of a generic observable $F(\omega_t)$ on `survived' trajectories of the reversed process as
\begin{align}
\mean{F}_{\tilde{\text{s}}} := \frac{1}{\tilde p^{\text{s}} (t)} \sum_{\omega_t \in \Omega_t } \chi(\tilde O(\omega_t)) \, F(\omega_t) \, P( \omega_t) ,
\end{align}
equation \eqref{psps} reads 
\begin{align}
p^\text{s}(t) &=\tilde p^\text{s}(t) \mean{e^{-\int_0^t dt' \dot \Sigma/k_{\text{B}}}}_{\tilde{\mathrm{s}}}
\ge \tilde p^{\text{s}}(t) e^{-  \frac{1}{k_\text{B}} \int_0^t dt' \langle  \dot \Sigma \rangle_{\tilde {\text{s}} }} ,
\label{psps2}
\end{align}
where we used Jensen's inequality in the last passage.

Since survival probabilities are positive and monotonically decreasing, one can define the instantaneous rates of the processes as
\begin{align}\label{rate}
&r(t):=- \frac{1}{ p^\text{s}(t)} \frac{d p^\text{s}}{d t}(t) , &&\tilde r(t):=- \frac{1}{  \tilde p^\text{s}(t)} \frac{d \tilde  p^\text{s}}{d t}(t).
\end{align} 
Since \eqref{psps2} holds for all times $t$, it gives the bound
\begin{align}\label{pbound2}
\frac{1}{k_{\text{B}}}\langle \dot \Sigma \rangle_{\tilde{\mathrm{s}}}(t')  \geq r(t')-\tilde r(t') .
 \end{align}

It is worth noticing the relation with previous results involving absolutely irreversible dynamics, in which some trajectories lack their time-reversed ones. In Ref.~\cite{mur14} the authors exponentiate also the indicator function $\chi$ which select absolutely continuous trajectories, namely,
\begin{align}\label{mur}
p^{\text{s}}(t) &= \sum_{\omega_t \in \Omega_t }e^{-\frac{1}{k_\text{B}} \int_0^t \dot \Sigma ( \omega_t) + \log \chi(O(\tilde \omega_t)) } P( \omega_t) .
\end{align}
Application of Jensen's inequality in \eqref{mur} would give a trivial bound on $p^\text{s}(t)$ because ${ \sum_{\omega_t}  \log \chi(O(\tilde \omega_t))P( \omega_t) = -\infty }$ due to the `absorbed' trajectories. This is explicitly stated in their last example where a Langevin dynamics with absorbing condition is considered.
In Refs.~\cite{fun15} and \cite{mur18} the authors do not exponentiate $\chi$ and so keep the sum over a subset of trajectories, similarly to our approach. Application of Jensen's inequality,  requires to normalize the average to 1, and so to introduce the survival probability of the reversed process, $\tilde p^\text{s}(t)$.
 It is unclear whether in Refs.~\cite{fun15} and \cite{mur18} Jensen's inequality is applied to the unnormalized average, which would be erroneous, or if the implicitly definition of average therein contains the normalization factor $\tilde p^\text{s}(t)$.  
 Hence, to the best of our knowledge, the modified integral fluctuation theorem and the bound in \eqref{psps} never appeared before in a clear explicit form.

We conclude our proof of the dissipation-time uncertainty relation introducing 3 hypotheses:
\begin{itemize}
\item[i)] Both processes (defined by $O$ and $\tilde O$, respectively) are rare. This means that their survival probabilities read $p^{\text{s}}(t)= e^{-r t}$ and $\tilde p^{\text{s}}(t)= e^{-\tilde r t}$. 
\item[ii)] The choice of $D$ is such that the rates satisfy $\tilde r \ll r$. Then, for times $t  \ll1/ \tilde r $, $\tilde p^{\text{s}}(t) \simeq 1$ and
$\mean{F}_{\tilde s}$ coincides with the unconstrained mean $\mean{F}:= \sum_{\omega_t \in \Omega_t  } F(\omega_t) P(\omega_t)$.
\item[iii)] The mean entropy production contains an entropy flow contribution which is extensive in time on the above mentioned timescale, namely,
\begin{align}
\int_0^t dt' \langle \dot \Sigma \rangle= \langle \dot S_\text{e} \rangle t.
\end{align}
\end{itemize}
Hence, \eqref{pbound2} turns into our final result $\langle \dot S_\text{e} \rangle \mathcal{T} \ge k_\text{B}$.

\section{Alternate derivation for integrated currents}

We consider a system in a steady state for which the fluctuation relation holds for the vector of time-averaged currents $J$ in the (long) time span $t$, 
\begin{align}\label{ft}
P(J)=P(-J) e^{t A \cdot J},
\end{align}
where $P(J)$ is the probability density of the current $J$ and $A$ is the vector of thermodynamic affinities (in units of $k_\text{B}$).
We define the process of realizing a scalar current larger than $J_i^*$, namely, we set the observable equal to the $i$th component of the vector $J$, $O=J_i$, and take $D= [J_i^*,+\infty)$.
The survival probability, i.e. the probability that the current $J_i^*$ was not yet realized in the time $t$, is
\begin{align}
p^\text{s}(t)&= \sum_{\omega_t \in \Omega_t } \Theta(J_i^*-J_i(\omega_t)) P(\omega_t) \\
&= \sum_{\omega_t \in \Omega_t } \Theta(J_i^*-J_i(\omega_t)) \underbrace{\int dJ \delta(J-J(\omega_t))}_{=1}P(\omega_t)\\
&= \int dJ \Theta(J_i^*-J_i) \underbrace{  \sum_{\omega_t} \delta(J-J(\omega_t)) P(\omega_t)}_{=:P(J)}\\
&= \int_{J_i<J_i^*} dJ P(J).
\end{align}
Using the fluctuation theorem \eqref{ft} we obtain
\begin{align}
p^\text{s}(t)&= \int_{ J_i < J_i^*} dJ P(-J) e^{t A \cdot J}\\
&=\int_{ J_i > -J_i^*} dJ P(J) e^{-t A \cdot J}  \\
&= \tilde p^{\text{s}}(t) \frac{\int_{ J_i > -J_i^*} dJ P(J) e^{-t A \cdot J} }{\int_{ J_i > -J_i^*} dJ P(J) } \ge \tilde p^{\text{s}}(t) \, e^{-t A \cdot  \int_{ J_i > -J_i^*} dJ J P(J)  } ,\label{intJ} 
\end{align}
where we changed variable to $\tilde J=- J$ and rename it as $\tilde J \to J$, and made use of Jensen's inequality. Here $\tilde p^{\text{s}}(t):= \int_{ J_i > -J_i^*} dJ P(J) $ is the probability that a current smaller than $-J^*_i$ is not realized in the time span $t$.

We then choose a threshold current $J^*_i>0$ sufficiently far from the mean value $\mean{J_i}>0$ (the choice of the sign is conventional) for the process to be rare, i.e. $p^{\text{s}}(t)=e^{-kt}$, but close enough so that $-J^*_i$ belongs to the negative tail of $P(J)$. This condition can always be realized away from equilibrium in the regime of weak fluctuations, where $\mean{J_i} \gg \sqrt{ \text{Var} {J_i}}$.
Under these conditions we can extend the integral over $J_i$ to $-\infty$ so that $\tilde p^{\text{s}}(t) =1$ and \eqref{intJ} becomes
\begin{align}
e^{-rt}=p^{\text{s}}(t)\ge  e^{-t A \cdot \mean{ J} }.
\end{align}
Since $A \cdot \mean{ J} = \langle \dot S_e \rangle / k_\text{B}$, is the mean entropy flow and $r=1/\mathcal{T}$ we retrieve our result $\langle \dot S_e \rangle \mathcal{T} \ge k_\text{B}$.

\section{Entropy flow and escape rate in drifted diffusion}

We consider within our theory the results of \cite{bre96, tel96} on the proportionality between escape rate and entropy production rate in large diffusive systems with constant drift. First, we reproduce the direct analytical calculation of the two quantities. Then, we show that their proportionality does not descends from the Eq. (8) of the main text, but instead it  corresponds to a saturated bound obtained by comparison with an auxiliary dynamics other then time-reversal.

We examine the Langevin dynamics 
\begin{align}\label{lang}
\dot x = f + \sqrt{2 /\beta}\xi,
\end{align}
where $f$ is a constant drift and $\xi$ a Gaussian white noise with zero mean and unit variance.
We impose absorbing conditions in $x=0$ and $x=L>0$ which amounts to consider the process
\begin{align}\label{processL}
&O=x_{t'}  && D= (-\infty, 0] \cup [L, \infty)
\end{align}
The associated state probability at time $t$ is given by the series $\rho_t(x)= \sum_{n=1}^\infty c_n e^{-r_n t} \exp(f x k_\text{B}\beta/2) \sin(n \pi x/L)$ with decay rates $r_n= f^2 \beta k_\text{B}/4 + n^2 \pi^2/(\beta k_\text{B} L^2)$ and coefficients $c_n$ depending on the initial conditions \cite{tel96}. In the limit of weak noise, i.e. $\beta \to \infty$, and/or large system size $L \to \infty$, all modes $n \neq 1$ are irrelevant and the probability decays with a single rate $r_1= f^2 \beta k_\text{B}/4= 1/\mathcal{T}$. This is 4 times smaller than the mean entropy flow associated to \eqref{lang} in absence of absorbing conditions, i.e. $\langle \dot S_\text{e}\rangle= f^2 k_\text{B} \beta$. Note that Ref.~\cite{bre96} incorrectly states the equality of escape rate and dissipation rate, which is due (in our notation) to the erroneous additional factors 2 and $1/2$ in the escape rate and entropy flow, respectively.

Even if the bound $\langle \dot S_e \rangle \mathcal{T} = 4 k_\text{B} \ge k_\text{B}$ is satisfied, our derivation does not apply  since the process \eqref{processL} is invariant under time-reversal, i.e. $p^\text{s}= \tilde p^{\text{s}}$.
However, the derivation can be  modified considering the auxiliary dynamics given by \eqref{lang} with $f=0$. The probability $P^\dagger(\omega_t)$ of the associated trajectories  can be used to write the survival probability of the process
\begin{align}
p^{\text{s}}(t)&= \sum_{\omega_t \in \Omega_t} \chi(O(\omega_t)) \frac{P(\omega_t)}{P^\dagger(\omega_t)} P^\dagger(\omega_t) \\
&=  \sum_{\omega_t \in \Omega_t} \chi(O(\omega_t)) e^{-\frac{1}{4 }f^2 \beta t  +\frac{1}{2} f \beta \int_0^t dt' \dot x }P^\dagger(\omega_t) \\
& \ge \tilde p^{\text{s}}(t) e^{-\frac{1}{4 }f^2 \beta k_\text{B} t  +\frac{1}{2} f \beta k_\text{B}  \int_0^t dt' \mean{\dot x}_{\tilde{\text{s}}} }  \label{pdagger}
\end{align}
where $\tilde p^{\text{s}}(t):=  \sum_{\omega_t \in \Omega_t} \chi(O(\omega_t)) P^\dagger(\omega_t)$ is the survival probability of the (auxiliary) process \eqref{processL} on the dynamics without drift, and $\mean{F}_{\tilde{\text{s}}} :=  \sum_{\omega_t \in \Omega_t} \chi(O(\omega_t)) F(\omega_t) P^\dagger(\omega_t) /\tilde p^{\text{s}}(t)$ is the normalized average over the same dynamics. In the limit $\beta \to \infty$ and/or $L \to \infty$ the processes are rare. In particular, the process without drift has rate $r_2 \ll r_1$ so that $\tilde p^{\text{s}}(t)=1$ and $
\mean{\dot x}_{\tilde{\text{s}}} =\mean{\dot x} =0$. Hence, equation \eqref{pdagger} yields
\begin{align}
p^{\text{s}}(t)= e^{-r_1 t} \ge  e^{-\frac{1}{4 }f^2 \beta k_\text{B} t }.
\end{align}
This is the system-specific bound $f^2 \beta  = \langle \dot S_\text{e} \rangle/ k_\text{B} \ge 4 r_1= 4/\mathcal{T} $ that we found above to be realized in equality.
\end{widetext}

 \end{document}